\documentclass{article} 
\usepackage{iclr2027_conference,times}

\usepackage{url}
\usepackage{xcolor}
\usepackage{etoc}         
\usepackage{booktabs}
\usepackage{enumitem}
\usepackage{tabularx}
\usepackage{graphicx}
\usepackage{letltxmacro}
\usepackage{hyperref}
\usepackage{amssymb}
\usepackage{listings}
\usepackage{amsmath}
\usepackage{algorithm}
\usepackage{algpseudocode}
\usepackage{subcaption}

\usepackage[most]{tcolorbox}
\tcbuselibrary{listings,breakable,skins}

\definecolor{ExampleHeader}{HTML}{8FB7D6}
\definecolor{ExampleFrame}{HTML}{D7E3EE}
\definecolor{ExampleBack}{HTML}{FAFCFE}

\definecolor{RaiseOver}{HTML}{B34E48}
\definecolor{RaiseUnder}{HTML}{416F9F}
\definecolor{RaiseLive}{HTML}{4F765F}
\definecolor{RaisePromptBack}{HTML}{FBFCFE}

\newtcblisting[
  auto counter,
  number within=section
]{raiseprompt}[4][]{
  enhanced,
  breakable,
  lines before break=5,
  colback=RaisePromptBack,
  colframe=#2,
  colbacktitle=#2,
  coltitle=white,
  fonttitle=\bfseries\scriptsize,
  title={Example~\thetcbcounter: #3},
  title after break={Example~\thetcbcounter: #3 (continued)},
  label={#4},
  boxrule=0.65pt,
  arc=1.5mm,
  left=6pt,
  right=6pt,
  top=6pt,
  bottom=6pt,
  before skip=8pt,
  after skip=8pt,
  listing only,
  listing options={
    basicstyle=\ttfamily\fontsize{7.2pt}{8.6pt}\selectfont,
    breaklines=true,
    breakatwhitespace=true,
    columns=fullflexible,
    keepspaces=true,
    showstringspaces=false,
    aboveskip=0pt,
    belowskip=0pt
  },
  #1
}

\usepackage[T1]{fontenc}
\usepackage[varqu]{zi4}

\newtcblisting[auto counter]{datasetblock}[3][]{
  enhanced,
  breakable,
  lines before break=6,
  colback=ExampleBack,
  colframe=ExampleFrame,
  colbacktitle=ExampleHeader,
  coltitle=white,
  fonttitle=\bfseries\small,
  title={Listing~\thetcbcounter: #2},
  title after break={Listing~\thetcbcounter: #2 (continued)},
  label={#3},
  before skip=4pt,
  after skip=4pt,
  boxrule=0.6pt,
  arc=2mm,
  left=2pt,
  right=2pt,
  top=2pt,
  bottom=2pt,
  listing only,
  listing options={
  basicstyle=\ttfamily\fontsize{6.5pt}{7.2pt}\selectfont,
  breaklines=true,
  breakatwhitespace=true,
  columns=fullflexible,
  keepspaces=true,
  showstringspaces=false,
  aboveskip=0pt,
  belowskip=0pt
  },
  #1
}

\makeatletter
\AtBeginDocument{%
  \def\hyper@natlinkbreak#1#2{#1}%
}
\makeatother

\definecolor{linkblue}{RGB}{40,60,200}  

\hypersetup{
  colorlinks=true,
  linkcolor=linkblue,
  citecolor=linkblue,
  urlcolor=linkblue,
  filecolor=linkblue,
}

\usepackage[utf8]{inputenc}
\DeclareUnicodeCharacter{00A0}{~}

\title{RAISE: Reinforcing Access Control Policy Synthesis in LLMs via Symbolic Evaluation}

\author{Yingming Zhou, Adarsh Vatsa, William Eiers\\
Stevens Institute of Technology\\
\texttt{\{yzhou136, avatsa, weiers\}@stevens.edu}}

\usepackage{colortbl}
\definecolor{cedarGain}{HTML}{0000FF} 
\definecolor{cedarLoss}{HTML}{FF0000} 
\definecolor{cedarHeader}{HTML}{EDF3F8}
\definecolor{cedarStripe}{HTML}{F2F2F2}
\newcommand{\cedarcell}[2]{%
  \makebox[2.6em][r]{#1}%
  \makebox[2.25em][l]{\raisebox{-0.25ex}{%
    \fontsize{6}{6}\selectfont #2}}}
    
\newcommand{\cedarplain}[1]{\cedarcell{#1}{}}
\newcommand{\cedarup}[2]{\cedarcell{#1}{\textcolor{cedarGain}{\ensuremath{\uparrow}#2}}}
\newcommand{\cedardown}[2]{\cedarcell{#1}{\textcolor{cedarLoss}{\ensuremath{\downarrow}#2}}}

\newcommand{\cedartablestyle}{%
  \fontsize{9}{10}\selectfont
  \setlength{\tabcolsep}{4pt}%
  \setlength{\extrarowheight}{0.6pt}%
  \renewcommand{\arraystretch}{1.0}%
  \setlength{\aboverulesep}{0pt}%
  \setlength{\belowrulesep}{0pt}}

\iclrfinalcopy 
\begin{document}
\etocdepthtag.toc{mtmain}   %

\maketitle
\fancyhead{} 

\begin{abstract}
Translating natural-language access-control requirements into policies
requires careful reasoning about permissions, constraints, and
exceptions, and even frontier LLMs often produce policies that violate
the intended authorization semantics. We construct CedarInstruct, to our
knowledge the first dataset that supports both training and semantic
evaluation for formally verifiable Cedar policy synthesis. It contains
5,800 scenarios across 44 domains and 1,408 representing a single
synthetic organization, each with a verified target policy and an
executable verification plan. On this data we introduce RAISE, which
trains policy synthesizers from formal verification in two stages,
verified supervised fine-tuning (SFT) followed by a reinforcement
learning (RL) stage that learns from verifier signal. We find that SFT
succeeds largely by letting models express authorization logic they
already have, since untrained models rarely write valid Cedar but often
reason correctly when they do. After SFT, how the verifier's information
is used matters more than how much of it is used. Of six RL
instantiations that consume progressively richer verifier signal, only
RAISE-OC improves meaningfully on SFT; it turns failed checks and
symbolic counterexamples into guided exploration and learns from the
result with off-context GRPO. With about 5.4K verified scenarios and
LoRA fine-tuning, RAISE-OC trains Qwen3.5-9B to surpass zero-shot GPT-6
Astra and Claude Opus 5 by 13.33 and 16.26 percentage points in semantic
success on held-out scenarios, and training transfers to the
independently constructed CedarBench.  The implementation is available at \url{https://github.com/Aizhouym/raise}.
\end{abstract}

\section{Introduction}
\label{sec:intro}

Access control policies decide which principals may perform which
actions on protected resources, and writing them precisely is hard.
Manual policy configuration is error-prone, and deployed authorization
often grants more access than intended \citep{xiao2025caspr,
cao2024stateful}. Translating natural-language requirements into
policies adds a further difficulty, because a policy must be both
syntactically valid and faithful to the requirement's authorization
semantics \citep{cheng2026say}. We study this problem in Cedar, a
declarative authorization language whose schemas and symbolic analysis
tools support validation, policy comparison, and counterexample
generation. \citep{cutler2024cedar} Appendix~\ref{app:cedar-example} shows the following policy in Cedar and a short guide to reading Cedar policies. Consider a requirement from our
dataset: finance managers may issue
payments only during business hours, and only for approved invoices
unless the request comes from the corporate network. A natural-looking
policy that grants the network exception as a separate rule is valid
Cedar, but it also lets managers on the corporate network issue payments
outside business hours, which the requirement forbids. Mistakes like
this, where an exception quietly overrides a restriction it was never
meant to lift, are easy to make and hard to spot by reading the policy.

Even frontier models find it hard to write a correct policy in a single
attempt. We check each generated policy against a \emph{verification
plan}, an executable set of symbolic checks that bound the access a
policy may grant and specify the access it must grant. Against these
plans, GPT-6 Astra and Claude Opus 5 produce valid Cedar for over 97\%
of our test scenarios but satisfy the full plan for only about a third
(Section~\ref{sec:results}). Verifier-guided repair can close part of
this gap \citep{vatsa2026autocedar}, but it needs the verifier, the plan,
and several model calls at deployment time. This paper asks whether a
small open model can instead be trained to generate correct policies
directly.

Answering that question requires verified demonstrations and executable
semantic criteria, so we first construct CedarInstruct, which contains
5,800 scenarios across 44 domains and 1,408 representing a single
synthetic organization, each pairing a requirement and a Cedar schema
with a verified target policy and a verification plan. Training on CedarInstruct follows a standard two-stage recipe. We first
fine-tune the model on the verified target policies (SFT), and then
continue with reinforcement learning (RL), rewarding each sampled policy
that passes verification. The two stages turn out to behave very
differently. SFT carries most of the improvement. Before training, many
policies are not valid and the smaller model's valid policies are rarely
correct, while after SFT nearly every policy is valid and more of them
are correct. SFT also plateaus quickly, since another epoch changes
nothing (Section~\ref{sec:why-sft}). RL, in contrast, has little to learn
from. With nearly every policy now valid, the remaining errors are
semantic, and for about a third of the inputs sampled during RL, every
sampled policy fails verification, so a pass-or-fail reward gives the
model no signal. Yet the verifier knows far more than pass or fail. It
can identify which property each policy violates and produce a concrete
request that witnesses the violation, and when we show these diagnostics
to the model, it repairs many of its failures (Section~\ref{sec:stall}).
The missing signal is there to be used.

We therefore introduce RAISE, which keeps this two-stage recipe but
builds the RL stage around the verifier's diagnostics rather than its
pass-or-fail verdict. We instantiate
the RL stage six ways, each adapted to Cedar synthesis and each consuming
more verifier information than the last, from binary outcomes up to
failed-check descriptions. The strongest instantiation, RAISE-OC, explores each verified failure in
its own guided context and learns from it with off-context GRPO
\citep{agrawal2026off}. It is the only instantiation that improves
meaningfully on SFT, even though two others receive the same
failed-check descriptions, which suggests that how verifier information
is used matters more than how much of it is used. The resulting 9B model
surpasses zero-shot frontier models on held-out scenarios, and training
transfers to the independently constructed CedarBench. Within a single
organization's fixed schema, training on about a thousand cases lets the
model solve nearly all new requirement combinations.

This paper makes three contributions. We construct \textbf{CedarInstruct}
(Section~\ref{sec:cedarinstruct}), to our knowledge the first dataset
supporting both training and semantic evaluation of formally verifiable
Cedar policy synthesis. We introduce \textbf{RAISE} (Section~\ref{sec:raise}), a
framework for learning from formal verification, instantiated six ways
along a ladder of verifier information, including RAISE-OC's
construction of guidance from verification. Finally, we analyze what
each training stage contributes (Sections~\ref{sec:setup}
and~\ref{sec:analysis}), showing that SFT makes policies valid and
plateaus quickly, that RL stalls where the verifier has most to say, and
that only verifier-guided exploration recovers that signal.

\section{Background}
\label{sec:background}
\label{sec:verification}

Two ideas underpin the rest of the paper, namely how we decide whether
a generated policy is correct and why standard reinforcement learning
can fail to improve a model that is already reasonably good.

\noindent\textbf{Policy correctness through symbolic verification.}
A Cedar request asks whether a principal may perform an action on a
resource in a given context. A schema $\Sigma$ declares the entity types,
attributes, actions, and context fields available, and a policy bundle
$P$ is a set of \texttt{permit} and \texttt{forbid} rules that authorizes
a request if and only if some \texttt{permit} rule applies and no
\texttt{forbid} rule does. Our task maps a requirement $D$ and a schema
$\Sigma$, written $q = (D, \Sigma)$, to a bundle $P$, and we judge $P$
by the set of requests it authorizes, written $\mathrm{sem}_\Sigma(P)$.

To decide correctness, we use boundary plans generated by AutoCedar
\citep{vatsa2026autocedar}. A plan $\Pi$ is a set of checks. Each check
$c$ pairs a reference policy $R_c$ with a request domain $D_c$ and has
one of three types. A \emph{ceiling} check ensures that no request
authorized by $P$ within $D_c$ falls outside the bound defined by $R_c$.
A \emph{floor} check requires $P$ to authorize every request authorized
by $R_c$ within $D_c$. A \emph{liveness} check requires $P$ to authorize
at least one such request, ensuring that the corresponding workflow
remains possible. Formally, the three check conditions are
\begin{equation}
\label{eq:check-conditions}
\begin{aligned}
\text{Ceiling:}\quad&
\mathrm{sem}_\Sigma(P) \cap D_c
\subseteq \mathrm{sem}_\Sigma(R_c),\\
\text{Floor:}\quad&
\mathrm{sem}_\Sigma(R_c) \cap D_c
\subseteq \mathrm{sem}_\Sigma(P),\\
\text{Liveness:}\quad&
\mathrm{sem}_\Sigma(P)
\cap \mathrm{sem}_\Sigma(R_c)
\cap D_c
\neq \varnothing.
\end{aligned}
\end{equation}

A symbolic checker, which we call the \emph{verifier}, decides these
conditions using Cedar's symbolic compiler (\texttt{cedar symcc}), which
reduces each check to an SMT query \citep{cutler2024cedar}. When a check
fails, the verifier returns its natural-language description and, for
ceiling and floor failures, a counterexample request whenever the solver
produces one. A policy is \emph{correct} if it parses, validates against
$\Sigma$, and satisfies every check, and we call the percentage of
generated policies that are correct \emph{semantic success}, our primary
metric. Correctness is therefore relative to the plan, which is used only
for training and evaluation and is never given to the model as input.

\label{sec:ocgrpo}
\noindent\textbf{The learning cliff and off-context guidance.}
Correctness also provides a natural reward. GRPO samples $K$ candidates
per input, rewards each with $r_k = 1$ if $P_k$ is correct and $0$
otherwise, and normalizes rewards within the group to obtain advantages
\citep{shao2024deepseekmath}. The normalization is also its weakness,
because when every candidate fails, all advantages vanish and the input
contributes no gradient. This is known as the \emph{learning cliff}
\citep{zhang2026scaf, agrawal2026off}, and several methods break it by
adding privileged guidance, such as hints or prefixes of a reference
solution, to the prompt so that hard inputs yield some successes
\citep{zhang2026scaf,qu2024recursive}.

Training on the guided prompt, however, optimizes a different objective
from the unguided one used at deployment. Off-context GRPO (OC-GRPO)
\citep{agrawal2026off} corrects this mismatch. It applies guidance $g$
only to inputs the policy cannot solve, samples under $(q, g)$, and
reweights each sampled token by
\begin{equation}
\label{eq:ocratio}
\rho_t(\theta) =
\frac{\pi_\theta(o_t \mid q, o_{<t})}
{\pi_{\theta_{\mathrm{old}}}(o_t \mid q, g, o_{<t})},
\end{equation}
which keeps the update unbiased for the unguided objective. Its analysis
adds two properties that matter later. The correction's variance grows
with the length of the guidance, which favors the shortest guidance that
breaks the cliff, and a guided success keeps credit only to the extent
that the model could have produced it unguided. What OC-GRPO leaves open
is where good guidance should come from, since its experiments use
prefixes of reference solutions.

\section{CedarInstruct}
\label{sec:cedarinstruct}

Learning from verification requires data in which every requirement comes
with a checkable definition of correct behavior. CedarInstruct provides this in two parts, 5,800 scenarios across 44 domains, which this section
describes and our main experiments use, and 1,408 scenarios representing
a single synthetic organization. 
Each scenario consists of a requirement $D$, a schema $\Sigma$, a
verification plan $\Pi$, reference policies, and a verified target
policy $P^\star$.

Each multi-domain scenario starts from a structured specification, a
machine-readable list of the actions the scenario governs and the rules
that apply to each. Specifications are sampled from hand-written domain
skeletons, which declare a domain's roles, resources, attributes, and
actions, and the 44 skeletons span 19 application areas. The rules
combine constraints such as separation of duty, lifecycle gates, and
overrides with conditions such as roles, thresholds, and network or time
restrictions, and the sampler rejects duplicates so that every scenario
is a distinct combination. GPT-5.5 then renders each specification as a
natural-language requirement, and AutoCedar \citep{vatsa2026autocedar}
builds the schema, plan, reference policies, and target policy. We keep a
scenario only if its policies validate and $P^\star$ satisfies every
check.

We split the multi-domain scenarios into 4,386 for SFT, 1,039 for RL,
and 375 for testing. No test requirement, schema, target policy, or plan
appears in training, but the test split draws on the same 44 domain
skeletons and the same construction pipeline, so it measures
in-distribution performance. To measure transfer, we also evaluate on
CedarBench \citep{vatsa2026autocedar}, whose scenarios were constructed
independently. Appendix~\ref{sec:dataset-construction} details the construction pipeline, sampling rules, and review procedures.

\section{RAISE: Learning from Verifier Signal}
\label{sec:raise}

\begin{figure}[t]
\centering
\includegraphics[width=\linewidth]{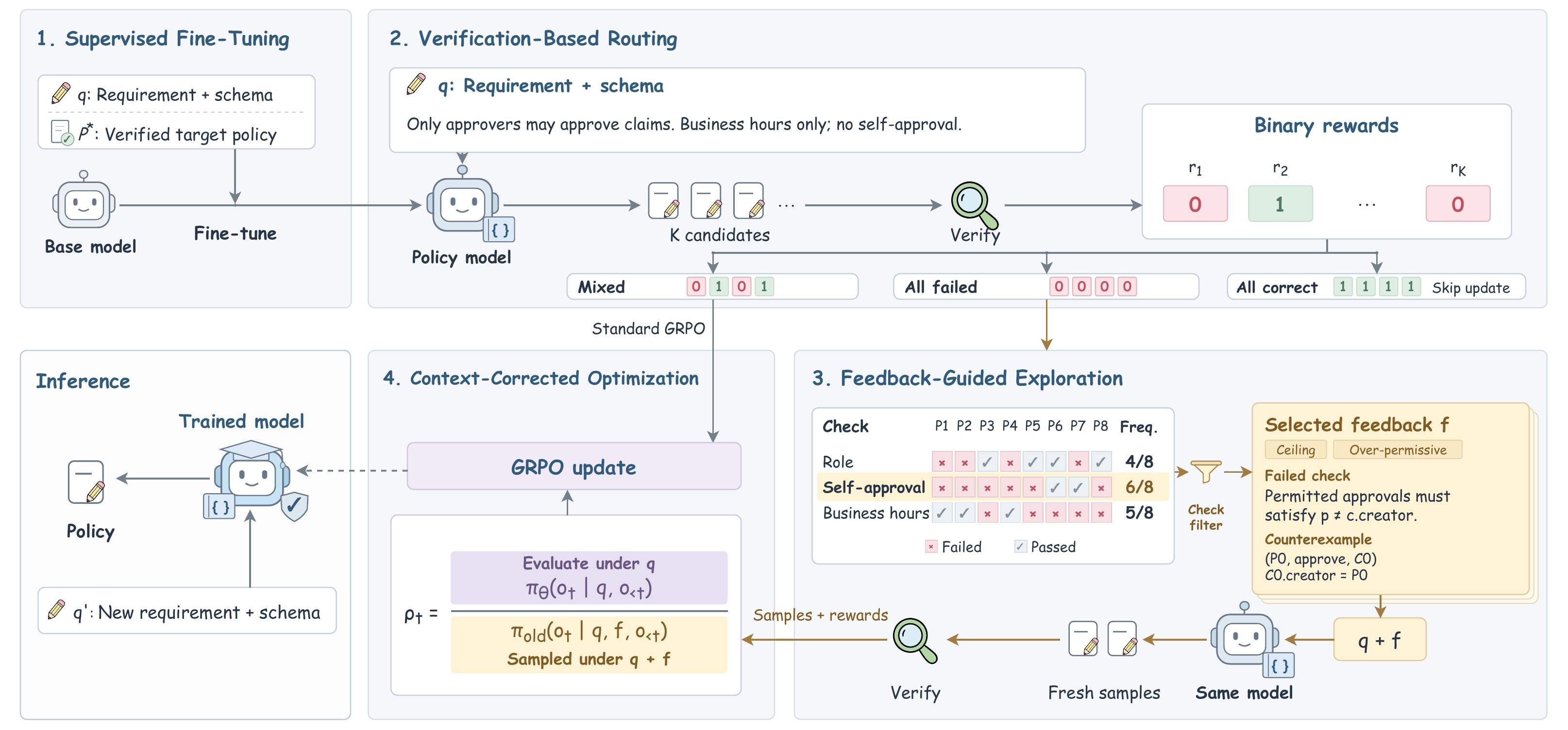}
\caption{Overview of RAISE. Verified SFT is followed by an RL stage that
learns from verifier signal. Shown is RAISE-OC, which routes all-failure
groups to verifier-guided exploration and learns from the guided samples
with an off-context update. At inference time, the model receives only
the requirement and schema.}
\label{fig:overview}
\end{figure}

CedarInstruct supplies verified policies to learn from, and the verifier
of Section~\ref{sec:verification} can judge any new policy the model
writes. RAISE uses both, training a policy synthesizer in two stages
(Figure~\ref{fig:overview}). The first teaches the model to write
verified policies, and the second continues training from the verifier's
feedback on the model's own attempts. In both stages, and at inference,
the model sees only the requirement and schema.

\subsection{Training Stages}
\label{sec:training-stages}

\noindent\textbf{Verified supervised fine-tuning.}
\label{sec:sft}
The first stage is supervised fine-tuning on verified target policies.
The model learns to generate $P^\star$ from $q$, and only the policy
tokens are supervised, with no intermediate reasoning. We hypothesize
that this generalizes rather than memorizes. \citet{lin2025debunk} trace
much of SFT's apparent failure to generalize to frozen prompts, which let
a model ignore its instructions, and find that SFT generalizes to new
instructions when each prompt declares its own vocabulary. CedarInstruct
has this property by construction, because every input declares its
vocabulary through its schema and a valid policy must use it.

\medskip
\noindent\textbf{Learning from verifier signal.}
\label{sec:rl-stage}
SFT still leaves many inputs unsolved, and the second stage learns from
the verifier on exactly those. 
Since no established method trains access-control synthesizers on sound verifier feedback, we adapt six verifier-guided training methods and order them by how much of the verifier's output each one uses
(Table~\ref{tab:instantiations}). 
They range from a binary pass-or-fail
reward, through the fraction and identity of the checks a candidate
passes, to the descriptions of the checks it fails. We ensure that all
six start from the same SFT checkpoint and see only $q$ at inference.

\begin{table}[t]
\centering
\caption{Six instantiations of the RAISE RL stage along the
verifier-feedback ladder.}
\label{tab:instantiations}
\begingroup
\cedartablestyle
\setlength{\tabcolsep}{5pt}
\renewcommand{\arraystretch}{1.06}
\begin{tabularx}{\linewidth}{
    >{\raggedright\arraybackslash}p{0.245\linewidth}
    >{\raggedright\arraybackslash}p{0.180\linewidth}
    >{\raggedright\arraybackslash}X}
\toprule
\rowcolor{cedarHeader}
\textbf{Method}
& \textbf{Verifier signal}
& \textbf{Training use} \\
\midrule
Binary Reward GRPO
& Plan outcome
& Binary reward; zero gradient for all-failure groups \\
Coverage Reward GRPO
& Check-pass fraction
& Graded reward with a degeneracy guard \\
Formal Dominance GRPO
& Passed-check set
& Set-dominance preferences within all-failure groups \\
Critique-GRPO
& Failure descriptions
& Verifier descriptions guide candidate refinement \\
SDPO
& Failure descriptions
& Self-teaching uses feedback and a successful sibling \\
\midrule
\textbf{RAISE-OC}
& Descriptions and counterexamples
& Separate guided contexts with off-context updates \\
\bottomrule
\end{tabularx}
\endgroup
\end{table}

\subsection{RAISE-OC}
\label{sec:raise-oc}

\noindent\textbf{Failure-based routing and guidance construction.}
The strongest instantiation, \textsc{RAISE-OC}, targets inputs for which
binary rewards provide no within-group learning signal. For each input
$q$, it samples $K$ candidates. Mixed-outcome groups receive a standard
GRPO update, while all-correct groups are skipped. For an all-failure
group, let $\mathcal{V}_k$ be the checks failed by candidate $k$,
$\mathcal{V}=\bigcup_{k=1}^{K}\mathcal{V}_k$, and
$m=\min(M,|\mathcal{V}|)$. Guidance is constructed as
\begin{equation}
h(c)=\sum_{k=1}^{K}\mathbf{1}[c\in\mathcal{V}_k],
\qquad
\mathcal{S}=\operatorname{Top}_{m}(\mathcal{V};h),
\qquad
\mathcal{B}=\{f_c=(d_c,x_c)\mid c\in\mathcal{S}\},
\label{eq:raise-guidance-construction}
\end{equation}
where $h(c)$ counts how many candidates fail check $c$, and $f_c$
contains its description $d_c$ and an available counterexample $x_c$.
Each $f_c\in\mathcal{B}$ defines a separate guided context. RAISE-OC
distributes $K$ fresh rollouts as evenly as possible across these
contexts. Restricting $M\leq\lfloor K/2\rfloor$ ensures at least two
rollouts per context. Every guided candidate is evaluated against the
complete verification plan, with advantages normalized within its
context.

\medskip
\noindent\textbf{Off-context optimization.}
Following OC-GRPO~\citep{agrawal2026off}, RAISE-OC uses the off-context
ratio $\rho^{\mathrm{OC}}_{c,\ell,t}(\theta)$ from
Equation~\ref{eq:ocratio} in a clipped surrogate:
\begin{equation}
\begin{aligned}
L_{\epsilon}(\rho,A)
&=
\min\!\left\{
\rho A,\,
\operatorname{clip}(\rho,1-\epsilon,1+\epsilon)A
\right\},\\
\mathcal{J}_{\mathrm{guided}}(\theta)
&=
\mathbb{E}_{\mathrm{guided}}
\!\left[
L_{\epsilon}\!\left(
\rho^{\mathrm{OC}}_{c,\ell,t}(\theta),
\widehat{A}_{c,\ell}
\right)
-\beta\mathcal{K}_{c,\ell,t}(\theta)
\right].
\end{aligned}
\label{eq:raise-oc-compact-objective}
\end{equation}
Here, $\mathbb{E}_{\mathrm{guided}}$ averages over guided inputs and
their candidates, with per-candidate token normalization as formalized
in Appendix~\ref{app:full-oc-objective}.
$\mathcal{K}_{c,\ell,t}$ is the token-level KL penalty to the fixed
reference policy, weighted by $\beta$. Because the numerator of
$\rho^{\mathrm{OC}}$ conditions only on $q$, guided samples update the
policy under the original input; no verifier feedback is therefore
needed at inference time. Appendix~\ref{app:full-oc-objective} gives the
full objective, Algorithm~\ref{alg:raise} the complete procedure, and
Appendix~\ref{sec:feedback-examples} a feedback example.

\medskip
\noindent\textbf{Design rationale.}
RAISE-OC derives guidance from the verification plan and the behavior of
the model's own candidates, without consulting the target policy
$P^\star$. Pooling failures prioritizes checks that block the most
candidates, while assigning each selected check a separate context
avoids combining unrelated diagnostics. Because every guided candidate
is evaluated against the complete plan, fixing one check while violating
another receives no reward. The off-context ratio and clipped surrogate
follow OC-GRPO; RAISE-OC contributes the verifier-based aggregation,
selection, and grouping of guidance.

\section{Experiments}
\label{sec:setup}

\subsection{Experimental Setup}

\textbf{Models and training.}
We now test whether this recipe delivers a small model that writes
correct policies, and whether the way RAISE-OC uses the verifier
matters. We train Qwen3.5-9B and Qwen3.8-27B with LoRA in verl
\citep{sheng2025hybridflow} on four NVIDIA H200 GPUs. Both backbones
undergo three primary SFT epochs, followed by two epochs of each RL
instantiation on the RL split, using a single training seed. For
Qwen3.5-9B, we additionally run a fourth SFT epoch to assess saturation;
all reported SFT results and subsequent RL runs use the epoch-3
checkpoint. RAISE-OC samples
$K = 8$ candidates per input and turns at most $M = 4$ failed checks
into guided contexts (Section~\ref{sec:raise-oc}).

\textbf{Baselines and prompts.}
For comparison, we evaluate GPT-6 Astra (Max) \citep{openai2026gpt6astra}
and Claude Opus 5 (Max) \citep{anthropic2026claudeopus5} with the zero-shot
prompt of Appendix~\ref{app:prompts}, and we also report GPT-5.5,
which helped construct the dataset, in
Appendix~\ref{app:construction-ref}. The untrained Qwen models
receive a structured Cedar prompt instead, because with the plain
prompt's brief guidance only 0.00\% and 0.27\% of their policies
are valid, as most generations fail parsing or schema validation
before any semantic check. The structured prompt adds syntax
patterns, schema-grounding constraints, and an output scaffold,
which makes it the stronger baseline.

\textbf{Evaluation protocol.}
We evaluate the final checkpoint of every training run, and no
prompt, hyperparameter, or checkpoint was chosen using the test
split. Every model generates one policy per scenario, greedily
for Qwen, with no verifier feedback.

\textbf{Evaluation metrics.}
Alongside semantic success we report syntax, the percentage of
valid policies, and two measures of partial correctness based on
individual checks. The macro score averages, over scenarios, the
fraction of each scenario's checks that the policy passes, so
every scenario counts equally. The micro score pools all checks
across scenarios and reports the fraction passed, so scenarios
with more checks count more. An invalid policy passes no checks.

\begin{table}[t]
\centering
\caption{Results on the CedarInstruct test split (\%). Every RL
instantiation starts from the SFT checkpoint, and arrows give the change
in semantic success relative to it. ``Among valid'' is semantic success
divided by syntax; for untrained models, valid policies are a selected
subset, so this rate describes the shift rather than isolating it.}
\label{tab:main}
\cedartablestyle
\begin{tabular}{lccccc}
\toprule
\rowcolor{cedarHeader}
Method & Syntax & Semantic & Among valid & Macro & Micro \\
\midrule
\rowcolor{cedarStripe}\multicolumn{6}{l}{\emph{Frontier models, zero-shot}} \\
GPT-6 Astra (Max) & 97.87 & \cedarplain{33.60} & 34.3 & 71.10 & 68.70 \\
Claude Opus 5 (Max) & 100.00 & \cedarplain{30.67} & 30.7 & 73.12 & 72.82 \\
\rowcolor{cedarStripe}\multicolumn{6}{l}{\emph{Qwen3.5-9B}} \\
Structured zero-shot & 45.07 & \cedarplain{2.67} & 5.9 & 23.51 & 22.25 \\
RAISE: SFT & 99.20 & \cedarplain{42.93} & 43.3 & 78.93 & 77.92 \\
\quad + Binary Reward GRPO & 98.93 & \cedarup{43.47}{0.54} & 43.9 & 78.94 & 77.92 \\
\quad + Coverage Reward GRPO & 98.40 & \cedardown{42.13}{0.80} & 42.8 & 77.56 & 76.36 \\
\quad + Formal Dominance GRPO & 97.33 & \cedarup{43.73}{0.80} & 44.9 & 77.99 & 77.12 \\
\quad + Critique-GRPO & 99.73 & \cedardown{42.67}{0.26} & 42.8 & 79.37 & 78.09 \\
\quad + SDPO & 97.07 & \cedarup{43.73}{0.80} & 45.0 & 78.22 & 76.81 \\
\quad + \textbf{RAISE-OC} & 99.20 & \cedarup{\textbf{46.93}}{4.00} & \textbf{47.3} & \textbf{80.78} & \textbf{79.34} \\
\rowcolor{cedarStripe}\multicolumn{6}{l}{\emph{Qwen3.8-27B}} \\
Structured zero-shot & 79.73 & \cedarplain{28.80} & 36.1 & 60.54 & 58.51 \\
RAISE: SFT & 99.73 & \cedarplain{46.40} & 46.5 & 81.02 & 80.03 \\
\quad + Binary Reward GRPO & 99.73 & \cedardown{44.80}{1.60} & 44.9 & 81.14 & 80.21 \\
\quad + Coverage Reward GRPO & 99.73 & \cedarup{46.93}{0.53} & 47.1 & \textbf{81.70} & \textbf{81.01} \\
\quad + Formal Dominance GRPO & 98.67 & \cedardown{43.47}{2.93} & 44.1 & 80.23 & 79.65 \\
\quad + Critique-GRPO & 99.73 & \cedardown{46.13}{0.27} & 46.3 & 80.44 & 79.31 \\
\quad + SDPO & 98.93 & \cedardown{45.60}{0.80} & 46.1 & 79.97 & 78.44 \\
\quad + \textbf{RAISE-OC} & 99.73 & \cedarup{\textbf{48.00}}{1.60} & \textbf{48.1} & 81.66 & 80.83 \\
\bottomrule
\end{tabular}
\end{table}

\subsection{Main Results}
\label{sec:results}
\label{sec:main-results}
In Table~\ref{tab:main}, five of the six RL instantiations improve semantic success by at most 0.80 points over SFT
on either backbone, and four of them fall below SFT on Qwen3.8-27B,
including Critique-GRPO and SDPO, which see the same failed-check
descriptions as RAISE-OC. RAISE-OC alone improves meaningfully, by 4.00
and 1.60 points, which corresponds to 15 and 6 additional scenarios
solved and a margin of 12 and 4 scenarios over the strongest other
instantiation (Appendix~\ref{app:significance}). Its validity is
unchanged, so the gains come from getting the semantics right. Scores
based on individual checks tell a less clear story, because passing more
checks does not guarantee a correct policy. For Qwen3.8-27B, Coverage Reward GRPO scores marginally higher than RAISE-OC
in macro and micro checks while solving fewer plans.
We also report, interestingly, that Claude Opus 5 passes more
individual checks than GPT-6 Astra but fewer complete plans for each scenario.

\label{sec:frontier}
Against frontier systems, both trained models come out ahead. With
RAISE-OC, Qwen3.5-9B surpasses GPT-6 Astra and Claude Opus 5 by 13.33 and
16.26 percentage points in semantic success, and Qwen3.8-27B by 14.40 and
17.33, without verifier feedback at inference, using LoRA and about 5.4K
verified scenarios. These test
scenarios are held out but in-distribution
(Section~\ref{sec:cedarinstruct}), and Section~\ref{sec:cedarbench}
tests transfer to independently constructed ones. The trained models
also run at lower per-scenario latency and estimated cost
(Appendix~\ref{app:efficiency}).

\section{Further Analysis}
\label{sec:analysis}

The main results raise two questions. Why does SFT account for most of
the gain, and why does RAISE-OC improve on it when other ways of using
the verifier do not? We take them in turn, and then ask how far the
results extend beyond the test split.

\subsection{What SFT and RAISE-OC Each Contribute}
\label{sec:why-sft}

Semantic success requires validity, and before training many policies
are not valid. Under structured zero-shot prompting, only 45.07\% of
Qwen3.5-9B's policies and 79.73\% of Qwen3.8-27B's are valid, and
correctness among the valid ones differs sharply between Qwen3.5-9B and
Qwen3.8-27B. Only 5.9\% of Qwen3.5-9B's valid policies satisfy the full
plan, whereas 36.1\% of Qwen3.8-27B's do, comparable to GPT-6 Astra
(34.3\%) and Claude Opus 5 (30.7\%) on the same measure
(Table~\ref{tab:main}). Our interpretation is that Qwen3.8-27B already
brings authorization reasoning from pretraining, likely from code in
other languages and other policy formats, but cannot yet express it
reliably in Cedar. The SFT data teaches that expression, meaning Cedar's
syntax and how its constructs interact, such as \texttt{permit} and
\texttt{forbid} rules, \texttt{when} and \texttt{unless} conditions, and
schema-typed attributes. After SFT, validity rises above 99\% for both
Qwen3.5-9B and Qwen3.8-27B, and correctness among valid policies rises to
43.3\% and 46.5\%. For Qwen3.8-27B, this is consistent with the view that
fine-tuning largely teaches models to express capabilities acquired in
pretraining \citep{zhou2023lima}, and it would help explain why about
4.4K demonstrations suffice. Qwen3.5-9B has less to unlock, since its
valid policies are rarely correct before training, so for Qwen3.5-9B,
SFT likely teaches reasoning as well as expression. The gap between
Qwen3.5-9B and Qwen3.8-27B among valid policies accordingly narrows from
30.2 points before training to 3.2 points after. Both frontier models achieve near-perfect syntactic validity (97.87\%
and 100.00\%), so their errors are unlikely to stem from unfamiliarity
with Cedar and instead reflect the difficulty of the authorization
reasoning itself.

If SFT mainly teaches a language, more of it should not help much, and it
does not. A further SFT epoch on the original 4,386 scenarios does not
improve Qwen3.5-9B, and a further epoch on the 1,039 RL-partition
scenarios, the same data RAISE-OC learns from, lowers its semantic
success by 5.60 points (Appendix~\ref{sec:implementation-details}). Further progress, therefore, needs a different training signal.

\begin{figure}[!t]
\centering
\setlength{\parskip}{0pt}

\begin{minipage}[t]{0.53\linewidth}
\vspace{0pt}
\centering

\includegraphics[width=\linewidth]{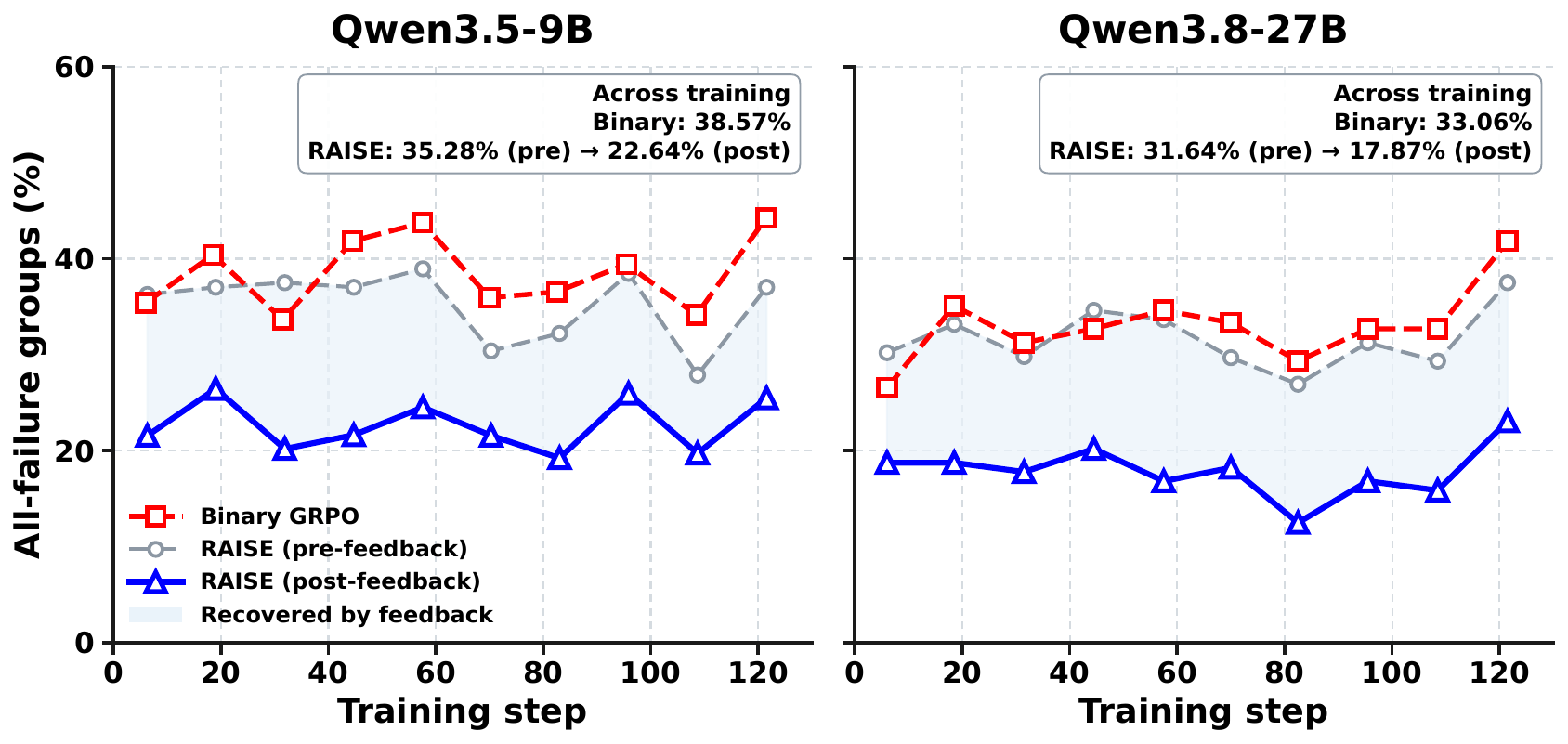}

\captionsetup{
  font=small,
  skip=3pt,
  justification=raggedright,
  singlelinecheck=false
}
\captionof{figure}{All-failure group rates for Binary GRPO and
RAISE-OC before and after verifier-guided exploration.
Shading indicates recovered learning signal.}
\label{fig:training}

\end{minipage}\hfill%
\begin{minipage}[t]{0.45\linewidth}
\vspace{0pt}
\centering

\captionsetup{
  font=small,
  skip=3pt,
  justification=raggedright,
  singlelinecheck=false
}
\captionof{table}{Component ablations: semantic success
(\%) on the test split.}
\label{tab:ablation}

\cedartablestyle
\setlength{\tabcolsep}{2pt}
\setlength{\extrarowheight}{0pt}
\renewcommand{\arraystretch}{1.0}

\begin{tabularx}{\linewidth}
  {@{}>{\raggedright\arraybackslash}Xcc@{}}
\toprule
\rowcolor{cedarHeader}
Configuration & 9B & 27B \\
\midrule

No SFT, binary RL
& \cedarplain{0.00}
& \cedarplain{27.20} \\

SFT
& \cedarplain{42.93}
& \cedarplain{46.40} \\

+ Binary GRPO
& \cedarup{43.47}{0.54}
& \cedardown{44.80}{1.60} \\

+ Guided, uncorrected
& \cedarup{44.27}{1.34}
& \cedarup{47.20}{0.80} \\

+ Guided, OC-GRPO
& \cedarup{\textbf{46.93}}{4.00}
& \cedarup{\textbf{48.00}}{1.60} \\

\bottomrule
\end{tabularx}

\par\vspace{2pt}
{\footnotesize\raggedright
\textit{Notes.}
9B: Qwen3.5-9B; 27B: Qwen3.8-27B.
Arrows show changes from SFT; the last row is RAISE-OC.
Syntax exceeds 99\% with SFT; without SFT, it is
0.00\% (9B) and 74.13\% (27B).
\par}

\end{minipage}
\end{figure}

\label{sec:stall}
\label{sec:ablation}
Once nearly every policy is valid, the remaining failures are semantic,
and under binary-reward GRPO, 38.57\% and 33.06\% of sampled groups
contain no correct candidate for the 9B and 27B models
(Figure~\ref{fig:training}), so they contribute nothing to learning.
These are the inputs where the model is most wrong, and they are also
where the verifier has the most to say, since beyond pass or fail it can
report the fraction and identity of the checks a candidate passes and,
for each failed check, a description of the violated property and often
a concrete counterexample. At inference time, giving the SFT model its
failed-check descriptions and counterexamples and asking it to repair its
policy raises semantic success by 21.07 points on Qwen3.5-9B, from 42.93\% to 64.00\%,
whereas a retry that reports only failure adds 2
(Appendix~\ref{app:repair}). The richest information is therefore directly useful to the model, and because its content rather than the extra attempt does most of the work, it serves, in the terms of
Section~\ref{sec:ocgrpo}, as guidance that raises the chance of success where the unguided model fails, without any reference solution.

Table~\ref{tab:ablation} isolates each component of RAISE-OC, with every
change measured relative to SFT. Without SFT, binary-reward GRPO fails on
both backbones. Qwen3.5-9B produces no valid policies, so every reward is
zero, and Qwen3.8-27B reaches 27.20\%, below the 28.80\% it achieves with
the structured prompt alone. After SFT, binary-reward GRPO changes
semantic success by $+0.54$ points on Qwen3.5-9B and $-1.60$ on
Qwen3.8-27B. With verifier-guided exploration and an uncorrected update,
the changes become $+1.34$ and $+0.80$. This variant samples under
feedback-augmented prompts but computes the loss under the original
input $q$, without the off-context correction. With the correction,
which gives RAISE-OC, they become $+4.00$ and $+1.60$, so the guidance
helps with either update and most with the correction. Within the
guidance, selecting checks by frequency, exploring each in its own
context, and pooling failures across candidates each outperform their
alternatives (Appendix~\ref{app:guidance-ablation}).
Figure~\ref{fig:training} shows where the gain comes from. Guided
exploration lowers RAISE-OC's all-failure rate from 35.28\% to 22.64\%
on Qwen3.5-9B and from 31.64\% to 17.87\% on Qwen3.8-27B, recovering a
learning signal for about 13\% of all sampled groups. Part of this
recovery comes from drawing $K$ additional candidates per all-failure
input, but the repair result above suggests that the diagnostics'
content, not the extra attempts, does most of the work.

\subsection{Beyond the Test Split}
\label{sec:cedarbench}

Our test scenarios are new, but they come from the same 44 domains as
the training data, with the same roles, resources, and actions. We
therefore evaluate on two further settings. The first is CedarBench,
whose scenarios were built independently of CedarInstruct. The second is
a single organization that keeps the same schema and actions while its
access rules change, tested on rule combinations unseen in training.

CedarBench~\citep{vatsa2026autocedar} was built independently of
CedarInstruct. To match CedarInstruct's input format, we convert CedarBench's inputs
into natural-language requirements using GPT-5.5 with the rendering
prompt of Listing~\ref{lst:render-prompt}, and we keep CedarBench's
original verification plans and scoring. None of its 221 scenarios is
used for training or model selection. Training transfers substantially
(Table~\ref{tab:cedarbench}). SFT raises semantic success from 0.00\% to
39.37\% on Qwen3.5-9B and from 1.36\% to 49.77\% on Qwen3.8-27B.
RAISE-OC is best on both backbones, reaching 43.44\% and 52.49\%, which
is 4.07 and 2.71 points above SFT, whereas binary-reward GRPO gains 3.17
and 0.00. On Qwen3.8-27B, RAISE-OC's gain comes despite lower syntax
validity than SFT, so it improves the correctness of valid policies
rather than their validity. Frontier models were not evaluated on
CedarBench, so the frontier comparison applies only to CedarInstruct.

\begin{table}[t]
\centering
\begin{minipage}[t]{0.585\linewidth}
\centering
\caption{CedarBench (221 scenarios, inputs converted to natural-language
requirements; \%) for Qwen3.5-9B and Qwen3.8-27B, excluded from training
and model selection. Arrows give the change in semantic success relative
to SFT.}
\label{tab:cedarbench}
\cedartablestyle
\resizebox{\linewidth}{!}{%
\begin{tabular}{lcccc}
\toprule
\rowcolor{cedarHeader}
& \multicolumn{2}{c}{9B} & \multicolumn{2}{c}{27B} \\
\rowcolor{cedarHeader}
Method & Syntax & Semantic & Syntax & Semantic \\
\midrule
Zero-shot & 0.00 & \cedarplain{0.00} & 1.81 & \cedarplain{1.36} \\
SFT & 64.71 & \cedarplain{39.37} & 72.40 & \cedarplain{49.77} \\
\quad + Binary GRPO & 66.97 & \cedarup{42.53}{3.17} & 72.40 & \cedarcell{49.77}{\textcolor{gray}{0.00}} \\
\quad + \textbf{RAISE-OC} & 66.52 & \cedarup{\textbf{43.44}}{4.07} & 68.78 & \cedarup{\textbf{52.49}}{2.71} \\
\bottomrule
\end{tabular}}
\end{minipage}\hfill
\begin{minipage}[t]{0.385\linewidth}
\centering
\caption{Semantic success (\%) of Qwen3.5-9B on 256 organizational test
cases. ``Before'' is the CedarInstruct SFT checkpoint; ``After'' is that
checkpoint after training on the 1,024 organizational training cases.}
\label{tab:organizational-adaptation}
\cedartablestyle
\resizebox{\linewidth}{!}{%
\begin{tabular}{lcc}
\toprule
\rowcolor{cedarHeader}
Input & Before & After \\
\midrule
Requirement + schema & 7.81 & \textbf{94.92} \\
\quad + plan & 58.20 & \textbf{96.88} \\
\bottomrule
\end{tabular}}
\end{minipage}
\end{table}
A deployed synthesizer faces a different kind of shift. An organization
typically keeps its schema and actions fixed while its requirements
change, and we study this setting with the 1,408 CedarInstruct scenarios
that represent a single synthetic organization. They share one
document-and-approval schema and six actions (read, edit, share, approve,
delete, and manage access), and we split them by rule combination into
1,024 training, 128 development, and 256 test cases, so every test case
is a combination of rules unseen in training. Starting from the SFT
checkpoint, we train Qwen3.5-9B for one epoch with a variant of the SDPO
instantiation whose self-teacher is conditioned on the verified reference
policy and on verified repair histories. We use this variant rather than
RAISE-OC because the setting is narrow. With the schema and actions
fixed, the same rules and checks recur across cases, much like a small
game, so we expect correct policies to be easier to reach and the
all-failure inputs that RAISE-OC targets to be rarer, while a teacher
that sees the verified reference provides a dense learning signal. One
model receives the requirement and schema, and a second also receives the
verification plan, as an organization that maintains its requirements as
checkable properties could provide. After training, the model given the
requirement and schema solves 243 of 256 test cases and the model also
given the plan solves 248 (Table~\ref{tab:organizational-adaptation}), so
modest organization-specific training lets Qwen3.5-9B handle new
requirement combinations almost without error. Before organization-specific training, the SFT checkpoint solves only
7.81\% of these test cases, against 42.93\% on the CedarInstruct test
split, so a single organization's rules can differ substantially from
our multi-domain data. Adding the plan raises its success to 58.20\%,
which suggests that the plan's atomic properties are useful scaffolding.

\section{Discussion}
\label{sec:discussion}

Four instantiations use more verifier information than a binary outcome,
and two use the same failed-check descriptions as RAISE-OC, yet only
RAISE-OC improves meaningfully over SFT. This suggests that the
difference lies in how the feedback is used. Critique-GRPO refines
candidates that have already failed, SDPO distills a feedback-conditioned
teacher into the student's own, mostly failing, candidates, and neither
corrects for the guided context. RAISE-OC instead samples fresh
candidates under short, targeted guidance and learns from them
off-context, which is where Section~\ref{sec:ocgrpo} predicts guidance is
most useful.

This guidance is exact because Cedar is deliberately not Turing-complete,
so whether a policy satisfies a check is decided by an SMT solver rather
than estimated with tests \citep{cutler2024cedar}. RAISE should therefore
transfer most directly to other policy and configuration languages that
can be analyzed the same way. The organizational study suggests a
practical deployment path, in which a broadly trained model is adapted to
one organization's fixed schema with about a thousand verified cases
(Section~\ref{sec:cedarbench}). Since SFT here is
answer-only, the atomic properties of each plan also offer a natural
reasoning scaffold, a direction supported by \citet{lin2025debunk}, who
find that chain-of-thought supervision helps on harder instances.

Our study has several limitations. Correctness is defined by generated
verification plans, whose fidelity symbolic checking cannot establish
and which we assess only through sampled human and LLM review. Because
the same pipeline defines the reward and the test grading, trained
models may partly learn its reading of ambiguous requirements.All results use a single training seed, so margins as small as 4 of 375
test scenarios (RAISE-OC over the strongest other instantiation on
Qwen3.8-27B) could partly reflect run-to-run variation. Guided
exploration also draws extra samples, and we did not compare verifier
guidance with reference-solution prefixes or LLM-written hints
(Appendix~\ref{app:limitations}).

\section{Related Work}
\label{sec:related-work}

Prior work applies LLMs to access-control policy generation through
prompting and fine-tuning \citep{lawal2024translating,
vatsa2025synthesizing, yang2025extraction, jayasundara2026agentvlm}.
AutoCedar instead uses symbolic verification for inference-time repair
\citep{vatsa2026autocedar}, whereas we use verification to train the
generator. In reinforcement learning with verifiable rewards
\citep{le2022coderl, guo2025deepseek}, GRPO estimates advantages from
within-group reward differences \citep{shao2024deepseekmath}.
Uninformative groups have been addressed through filtering and
resampling \citep{yu2026dapo}, adaptive task difficulty
\citep{zeng2025rlve}, and hierarchical hints \citep{zhang2026scaf}.
Feedback has supported both inference-time refinement
\citep{madaan2023self, gou2024critic} and training in Critique-GRPO and
SDPO \citep{zhang2025critique, hubotter2026reinforcement}, which we adapt
as RAISE instantiations. Closest to RAISE-OC, adaptive OC-GRPO retries
failed prompts with progressively stronger tutor-generated hints
\citep{agrawal2026off}. RAISE-OC instead derives guidance from formal
verification of the model's own candidates, pools failures across
candidates, and explores each selected property in its own context.
Appendix~\ref{app:related-work} provides a fuller review.

\section{Conclusion}
\label{sec:conclusion}

Small open models can learn to translate access-control requirements
into verified Cedar policies better than zero-shot frontier models, from
a few thousand verified scenarios and without verifier feedback at
inference. Verified SFT does most of the work, making nearly every policy
valid and more often correct, but it plateaus, and more supervised data
does not help. Beyond that point, the verifier's diagnostics supply the
missing signal, and how they are used decides whether they help. Of six
ways of learning from verification, only exploring each verified failure
in its own guided context and learning from it off-context improves
meaningfully on SFT. Training transfers to independently built
scenarios, and a synthesizer adapts to a single organization's
requirements from about a thousand examples.

\section*{AI Use Statement}
The authors used large language models (LLMs) to assist with manuscript
polishing and the construction of CedarInstruct. Specifically, LLMs were
used to render synthetic access-control requirements from structured
specifications and to assist with faithfulness review during dataset
construction. Language polishing focused on proofreading, grammatical
correction, and stylistic refinement. The authors conceived the research
ideas, designed the methodology, and conducted the experimental
evaluation and analysis.

\section*{Ethics Statement}
This research follows established research ethics principles. Our objective is to
improve the reliability of access control policy synthesis by
integrating formal verification with language model training. We
acknowledge the security risks of incorrectly generated policies,
including unintended permissions and denial of legitimate access. Our
symbolic verification guarantees are limited to the specified
verification plans. CedarInstruct is constructed from synthetic
scenarios, without real-world personal or sensitive authorization data.
We encourage deploying generated policies with independent validation
and human oversight.

\section*{Reproducibility Statement}
The supplementary material contains the complete source code, including
the implementation of RAISE and all six RL-stage instantiations, the
CedarInstruct construction and verification pipeline, and all training
and evaluation scripts. The implementation is available at \url{https://anonymous.4open.science/r/raise-424A/}.

\bibliography{iclr2027_conference}
\bibliographystyle{iclr2027_conference}

\clearpage

\appendix

\etocdepthtag.toc{mtappendix}
\etocsettagdepth{mtmain}{none}          
\etocsettagdepth{mtappendix}{subsection} 
\etocsettocstyle{\section*{Table of Contents}\vspace{0.6em}}{}
\tableofcontents
\clearpage

\section{Comprehensive Review of Related Work}
\label{app:related-work}

\paragraph{Access-Control Policy Synthesis and Verification.} Earlier work on natural-language policy authoring extracted access-control rules from textual requirements~\citep{xiao2012automated,slankas2014relation} and constructed authorization models from policy statements~\citep{abdelgawad2023synthesizing}. Recent studies investigate LLM prompting for policy extraction and synthesis, including whether generated policies preserve the intended authorization behavior~\citep{lawal2024translating,vatsa2025synthesizing,vatsa2025exploring}. Beyond prompting, \citet{yang2025extraction} fine-tune CodeT5 on GPT-4-generated data to extract ABAC rules from natural-language text. AGentVLM uses small open-source language models for structured policy extraction and provides verification feedback to support policy revision~\citep{jayasundara2026agentvlm}. Formal analysis complements policy generation by checking authorization behavior against specified properties. Zelkova and Cedar support symbolic reasoning about policies, including comparisons of the requests they allow~\citep{backes2018semantic,cutler2024cedar}. Verification also supports policy modification. Restricter tightens Cedar policies against access logs, while CloudFix combines formal fault localization with LLM-generated repairs checked by SMT solvers~\citep{wu2026automatically,hall2026cloudfix}. AutoCedar constructs reviewed schemas and boundary plans specifying required permissions, upper bounds on permitted access, and liveness conditions, and uses symbolic verification to guide policy generation and repair~\citep{vatsa2026autocedar}. Our focus is on using the resulting verification feedback to train the generator, rather than solely to assess and repair individual candidate policies.

\paragraph{Reinforcement Learning with Verifiable Rewards.} Verification results can serve as training rewards for policy generation. This follows the broader reinforcement learning with verifiable rewards (RLVR) paradigm, which trains language models using automatically assessed outcomes in mathematical reasoning and code generation~\citep{guo2025deepseek,team2025kimi}. Earlier code-generation methods use execution outcomes, learned critics, fine-grained unit-test feedback, and code-completion curricula to guide learning~\citep{le2022coderl,liu2023rltf,dou2024stepcoder}. GRPO estimates advantages relative to responses sampled for the same prompt, avoiding a separately trained value model~\citep{shao2024deepseekmath}. Subsequent methods and analyses examine how reward normalization and policy regularization affect optimization~\citep{liu2025understanding,mroueh2026reinforcement}. Sampling becomes particularly important when rewards offer little variation within a group. In GRPO with binary outcome rewards, groups containing only successful or only unsuccessful responses yield zero group-relative advantages. DAPO maintains informative training batches by filtering such groups and replenishing them through additional sampling~\citep{yu2026dapo}, while RLVE adapts the difficulty of procedurally generated tasks to the evolving policy~\citep{zeng2025rlve}. For policy generation, symbolic verification can identify violated authorization properties and, when available, return concrete requests that witness those violations. Descriptions of the violated properties and the corresponding counterexamples can guide subsequent policy generation, but this information is not conveyed by an aggregate correctness reward alone.

\paragraph{Feedback-Guided Training of LLMs.} Beyond scalar rewards, natural-language critiques and tool feedback have been used to guide iterative refinement without updating model parameters~\citep{madaan2023self,gou2024critic}. Training-based approaches teach models to revise their outputs through supervised learning and multi-turn reinforcement learning~\citep{qu2024recursive,kumar2025training,jiao2026thinktwice}. Related methods jointly train solution generation and correctness assessment using verifiable outcomes~\citep{liu2026trust,ruan2026critique}. More directly related to learning from assisted generations, Critique-GRPO uses natural-language critiques to generate refinements and optimizes both initial responses and refinements during online training~\citep{zhang2025critique}. SDPO uses a feedback-conditioned version of the model as a self-teacher and distills its next-token distributions into the policy conditioned on the original prompt~\citep{hubotter2026reinforcement}. Learning from assisted generations also requires accounting for how the training responses are obtained. LUFFY combines off-policy demonstrations with on-policy rollouts and uses policy shaping to balance imitation and exploration~\citep{yan2026learning}. OC-GRPO instead studies a mismatch in conditioning, generating rollouts with privileged guidance while optimizing an importance-corrected objective under the original prompt~\citep{agrawal2026off}. Our work investigates how structured evidence from formal verification can be used during training to improve the semantic correctness of generated access-control policies.
\newpage

\section{Formal Definitions and Training Details}
\label{app:formal}

This appendix gives the formal definitions behind Sections~\ref{sec:verification} and~\ref{sec:raise}: verification, the SFT and GRPO objectives, RAISE-OC's guidance construction and objective, an example, and the complete procedure.

\subsection{Training Procedure}

Algorithm~\ref{alg:raise} gives the complete procedure.

\begin{algorithm}[t]
\caption{RAISE with the RAISE-OC instantiation}
\label{alg:raise}
\footnotesize
\begin{algorithmic}[1]
\Require Initial parameters $\theta_0$; partitions
$\mathcal E_{\mathrm{SFT}},\mathcal E_{\mathrm{RL}}$; rollout budget
$K\ge2$; feedback limit $1\le M\le\lfloor K/2\rfloor$; RL epochs
$N_{\mathrm{RL}}$; optimization steps $U$; $\epsilon,\delta,\beta$
\Ensure A policy $\pi_\theta(\cdot\mid D,\Sigma)$ that needs no verifier
feedback at inference
\State $\theta\gets\operatorname{SFT}(\theta_0,\mathcal E_{\mathrm{SFT}})$
\Comment{Eq.~\ref{eq:sft-objective}}
\For{$e=1,\ldots,N_{\mathrm{RL}}$}
  \For{each minibatch $\mathcal Q\subset\mathcal E_{\mathrm{RL}}$}
    \State $\theta_{\mathrm{old}}\gets\theta$; $\mathcal G\gets[\,]$
    \For{each $(D,\Sigma,\Pi,\mathcal R)\in\mathcal Q$}
      \State $q\gets(D,\Sigma)$; sample
      $P_k\sim\pi_{\theta_{\mathrm{old}}}(\cdot\mid q)$,
      $k=1,\ldots,K$
      \State $(r_k,z_k)\gets\operatorname{Verify}(P_k;\Sigma,\Pi,\mathcal R)$
      \If{$\sum_k r_k=K$}
        \State \textbf{continue} \Comment{all correct: skip}
      \ElsIf{$\sum_k r_k>0$}
        \State $\Gamma_q\gets\{(q,\{(P_k,r_k)\}_{k=1}^{K})\}$
        \Comment{mixed: standard GRPO}
      \Else
        \State $V\gets\bigcup_k V(P_k)$;
        $h(c)\gets|\{k:c\in V(P_k)\}|$
        \If{$V=\varnothing$}
          \State $\mathcal B\gets\{f_{\mathrm{syn}}\}$
        \Else
          \State $J\gets$ the $\min(M,|V|)$ checks with largest $h(c)$;
          $\mathcal B\gets\{f_c=(d_c,x_c)\}_{c\in J}$
        \EndIf
        \State Choose $n_c\ge2$ as evenly as possible with
        $\sum_{f_c\in\mathcal B}n_c=K$; $\Gamma_q\gets[\,]$
        \For{each $f_c\in\mathcal B$}
          \State Sample
          $\widetilde P_{c,\ell}\sim
          \pi_{\theta_{\mathrm{old}}}(\cdot\mid q,f_c)$,
          $\ell=1,\ldots,n_c$;
          $\widetilde r_{c,\ell}\gets\operatorname{Verify}(\widetilde P_{c,\ell})$
          \State Append $((q,f_c),\{(\widetilde P_{c,\ell},
          \widetilde r_{c,\ell})\}_\ell)$ to $\Gamma_q$
        \EndFor
      \EndIf
      \State Append $(q,\Gamma_q)$ to $\mathcal G$
    \EndFor
    \If{$\mathcal G\neq[\,]$}
      \State Compute group-wise advantages and frozen behavior
      probabilities \Comment{Eq.~\ref{eq:full-oc-ratio}}
      \For{$u=1,\ldots,U$}
        \State $\theta\gets\theta+\eta\nabla_\theta
        \mathcal J_{\mathcal G}(\theta)$
        \Comment{Eq.~\ref{eq:full-oc-objective}}
      \EndFor
    \EndIf
  \EndFor
\EndFor
\State \Return $\pi_\theta$
\end{algorithmic}
\end{algorithm}

\subsection{Verification}

Let $\mathcal X_\Sigma$ denote the schema-conforming authorization inputs, each a request with the entity data needed to evaluate it. $\operatorname{Valid}_\Sigma(P)$ holds when $P$ parses and every policy in it validates against $\Sigma$; for a valid bundle, $\operatorname{sem}_\Sigma(P)=\{x\in\mathcal X_\Sigma\mid P\text{ authorizes }x\}$. A check $c\in\Pi$ has a type $\tau_c\in\{\mathrm{ceil},\mathrm{floor},\mathrm{live}\}$, a reference policy $R_c$, and a domain $D_c\subseteq\mathcal X_\Sigma$ given by its declared principal, action, and resource scope. A valid candidate satisfies $c$, written $P\models_\Sigma c$, when
\begin{equation}
P\models_\Sigma c \Longleftrightarrow
\begin{cases}
\operatorname{sem}_\Sigma(P)\cap D_c
  \subseteq \operatorname{sem}_\Sigma(R_c), & \tau_c=\mathrm{ceil},\\
\operatorname{sem}_\Sigma(R_c)\cap D_c
  \subseteq \operatorname{sem}_\Sigma(P), & \tau_c=\mathrm{floor},\\
\operatorname{sem}_\Sigma(P)\cap\operatorname{sem}_\Sigma(R_c)\cap D_c
  \neq\varnothing, & \tau_c=\mathrm{live}.
\end{cases}
\label{eq:formal-check}
\end{equation}
A ceiling failure has a witness in $\operatorname{sem}_\Sigma(P)\cap D_c\setminus \operatorname{sem}_\Sigma(R_c)$ and a floor failure a witness in $\operatorname{sem}_\Sigma(R_c)\cap D_c\setminus \operatorname{sem}_\Sigma(P)$; a liveness failure means the region contains no authorized input, so it has no witness. The diagnostic record of $P$ consists of its validity, the set $V(P)\subseteq\Pi$ of failed checks ($\varnothing$ when $P$ is invalid), and for each $c\in V(P)$ the description $d_c$ and, when the solver returns one, a counterexample $x_c$. The reward is
\begin{equation}
r(P)=\mathbb I\!\left[
\operatorname{Valid}_\Sigma(P)\land
P\models_\Sigma c\ \forall c\in\Pi\right].
\label{eq:formal-reward}
\end{equation}

\subsection{Verified SFT Objective}

With $o_i^\star=(o_{i,1}^\star,\ldots,o_{i,T_i}^\star)$ the tokens of the target policy for input $q_i$, SFT minimizes
\begin{equation}
\mathcal L_{\mathrm{SFT}}(\theta)
=-\frac{1}{\sum_{i=1}^{N}T_i}
\sum_{i=1}^{N}\sum_{t=1}^{T_i}
\log\pi_\theta\!\left(o_{i,t}^\star\mid q_i,o_{i,<t}^\star\right),
\label{eq:sft-objective}
\end{equation}
so only policy tokens contribute, weighted equally without per-example length normalization.

\subsection{Group Relative Policy Optimization}

Given $q$, the policy $\pi_{\theta_{\mathrm{old}}}$ samples $K$ responses $o_1,\ldots,o_K$ with rewards $r_1,\ldots,r_K$, and every token of $o_k$ receives the advantage $\widehat A_k=(r_k-\bar r)/(s+\delta)$, where $\bar r$ and $s$ are the mean and sample standard deviation of the group's rewards and $\delta>0$ ensures stability \citep{shao2024deepseekmath}. With the per-token ratio $\rho_{k,t}(\theta)=\pi_\theta(o_{k,t}\mid q,o_{k,<t})/ \pi_{\theta_{\mathrm{old}}}(o_{k,t}\mid q,o_{k,<t})$, GRPO maximizes
\begin{equation}
\mathcal J_{\mathrm{GRPO}}(\theta)=
\mathbb E\!\left[
\frac{1}{K}\sum_{k=1}^{K}\frac{1}{|o_k|}
\sum_{t=1}^{|o_k|}
\left(
\min\!\left(\rho_{k,t}\widehat A_k,
\operatorname{clip}(\rho_{k,t},1-\epsilon,1+\epsilon)\widehat A_k\right)
-\beta\operatorname{KL}_{k,t}
\right)\right],
\label{eq:formal-grpo}
\end{equation}
where $\operatorname{KL}_{k,t}=D_{\mathrm{KL}}( \pi_\theta(\cdot\mid q,o_{k,<t})\Vert \pi_{\mathrm{ref}}(\cdot\mid q,o_{k,<t}))$ and $\pi_{\mathrm{ref}}$ is the fixed SFT policy.

\subsection{RAISE-OC's Guidance Construction}

Given $K$ candidates with rewards $r_k$ and $n_+=\sum_k r_k$ successes, RAISE-OC routes each group as in OC-GRPO \citep{agrawal2026off}:
\begin{equation}
\operatorname{Route}(n_+)=
\begin{cases}
\text{standard GRPO update}, & 0<n_+<K,\\
\text{skip}, & n_+=K,\\
\text{verifier-guided exploration}, & n_+=0.
\end{cases}
\label{eq:formal-route}
\end{equation}
For an all-failure group, let $V=\bigcup_k V(P_k)$ be the distinct failed checks and $h(c)=|\{k:c\in V(P_k)\}|$ the number of candidates failing $c$; invalid candidates contribute no checks. RAISE-OC selects the set $J$ of the $\min(M,|V|)$ checks with the largest $h(c)$, where $1\le M\le\lfloor K/2\rfloor$. Each selected check yields a feedback item $f_c=(d_c,x_c)$, omitting $x_c$ when no counterexample is available. If no candidate is valid, so that $V=\varnothing$, RAISE-OC uses a single syntax-correction instruction $f_{\mathrm{syn}}$. The feedback set is $\mathcal B=\{f_c\}_{c\in J}$, or $\{f_{\mathrm{syn}}\}$ when $J=\varnothing$.

Each $f_c\in\mathcal B$ receives $n_c\ge2$ rollouts, as evenly as possible, with $\sum_c n_c=K$, and RAISE-OC samples
\begin{equation}
\widetilde P_{c,\ell}\sim
\pi_{\theta_{\mathrm{old}}}(\cdot\mid q,f_c),
\qquad \ell=1,\ldots,n_c.
\label{eq:guided-sampling}
\end{equation}
Each guided candidate receives the full-plan reward of Eq.~\ref{eq:formal-reward}, and advantages are normalized within the group of candidates that share $f_c$:
\begin{equation}
\widehat A_{c,\ell}=\frac{\widetilde r_{c,\ell}-\bar r_c}{s_c+\delta},
\qquad
\bar r_c=\frac{1}{n_c}\sum_{\ell=1}^{n_c}\widetilde r_{c,\ell},
\label{eq:guided-advantages}
\end{equation}
where $s_c$ is the sample standard deviation of the group's rewards. The initial $K$ candidates serve only for routing and guidance, so an all-failure input costs $2K$ generations, of which the $K$ guided ones enter optimization.

\subsection{When a Guided Group Yields a Learning Signal}

A group is outcome-informative when at least one advantage is nonzero; with binary rewards, exactly when it contains both a success and a failure. If candidates sampled under $(q,f_c)$ are independent with common success probability $p_c$, the group is outcome-informative with probability
\begin{equation}
g_{n_c}(p_c)=1-(1-p_c)^{n_c}-p_c^{n_c},
\label{eq:informative-probability}
\end{equation}
since the number of successes is $\operatorname{Binomial}(n_c,p_c)$. Two consequences follow. Guidance helps only if it raises $p_c$ above zero, which inference-time repair suggests verifier feedback does (Appendix~\ref{app:repair}). And when four checks are selected, each group has $n_c=2$ and is informative with probability $2p_c(1-p_c)\le1/2$, so exploration creates opportunities for learning signal rather than guaranteeing it.

\subsection{Full Off-Context Objective}
\label{app:full-oc-objective}

For a retained input $q$, let $\Gamma_q$ be its groups. A mixed-reward input has one group of $K$ candidates with behavior context $b_g=q$; an all-failure input has one group per feedback item, with $b_g=(q,f_c)$ and $n_g=n_c$ candidates, so that $\sum_{g\in\Gamma_q}n_g=K$. For each sampled token we freeze the behavior probability and define
\begin{equation}
\beta_{g,\ell,t}=\pi_{\theta_{\mathrm{old}}}
(o_{g,\ell,t}\mid b_g,o_{g,\ell,<t}),
\qquad
\rho_{g,\ell,t}(\theta)=
\frac{\pi_\theta(o_{g,\ell,t}\mid q,o_{g,\ell,<t})}
{\beta_{g,\ell,t}}.
\label{eq:full-oc-ratio}
\end{equation}
When $b_g=q$ this is the GRPO ratio; when $b_g=(q,f_c)$ it is the off-context ratio of Eq.~\ref{eq:ocratio}. With $\ell_\epsilon(\rho,A)=\min(\rho A, \operatorname{clip}(\rho,1-\epsilon,1+\epsilon)A)$ and group-normalized advantages $\widehat A_{g,\ell}$, each candidate contributes
\begin{equation}
\mathcal J_{g,\ell}(\theta)=
\frac{1}{|o_{g,\ell}|}\sum_{t=1}^{|o_{g,\ell}|}
\left[\ell_\epsilon\!\left(\rho_{g,\ell,t}(\theta),
\widehat A_{g,\ell}\right)-\beta\operatorname{KL}_{g,\ell,t}(\theta)\right],
\label{eq:full-oc-candidate}
\end{equation}
with the KL term evaluated under $q$. For a minibatch $\mathcal Q$ of retained inputs, RAISE-OC maximizes
\begin{equation}
\mathcal J_{\mathcal Q}(\theta)=
\frac{1}{|\mathcal Q|}\sum_{q\in\mathcal Q}
\frac{1}{K}\sum_{g\in\Gamma_q}\sum_{\ell=1}^{n_g}
\mathcal J_{g,\ell}(\theta).
\label{eq:full-oc-objective}
\end{equation}
Each retained input contributes exactly $K$ optimized candidates. For mixed-reward inputs the objective reduces to GRPO; for all-failure inputs it averages separately normalized groups without mixing their advantages. Because the numerator is always evaluated under $q$, every active term updates the policy conditioned on the original input, though advantages may vanish, clipping may suppress terms, and token-level gradients may cancel. The behavior probability must include any sampling transformation such as temperature, and the result is a clipped token-level surrogate in the style of OC-GRPO, not an unbiased sequence-level estimator.

\subsection{Illustrative Verifier Feedback}
\label{sec:feedback-examples}

To illustrate the feedback supplied during guided exploration, we construct controlled failure variants of the verified policy for scenario \texttt{invoice\_approval-issuePayment-T3-045}. These variants are used only to expose representative verifier outputs; they are not presented as historical model generations. We show the failed-check descriptions and the relevant projection of each symbolic counterexample. Solver-assigned fields unrelated to the failed property are omitted for readability.

\begin{datasetblock}[breakable=false]
{RAISE feedback suffix}
{lst:raise-feedback-template}
A verifier found your earlier attempts violated this requirement;
make sure your policy satisfies it:
- {FAILED_CHECK_DESCRIPTION}
  concrete case your earlier attempt decided wrongly:
  {REDUCED_COUNTEREXAMPLE}
\end{datasetblock}

The suffix is appended to the original schema--requirement prompt;
the earlier failed policy itself is not included. Each selected check
defines a separate feedback-conditioned rollout subgroup. The following
examples come from scenario
\texttt{invoice\_approval-issuePayment-T3-045}. For readability, the
symbolic witnesses are projected onto fields relevant to the failed
check; ellipses denote solver-assigned fields that are omitted only from
the presentation.

\begin{raiseprompt}
{RaiseOver}
{Over-permissive policy: business-hours ceiling violation}
{lst:raise-overpermissive-hours}
A verifier found your earlier attempts violated this requirement;
make sure your policy satisfies it:
- A finance manager must not be permitted to issue a payment
  outside business hours.
  concrete case your earlier attempt decided wrongly:
  principal: Person::"P0"
  action: Action::"issuePayment"
  resource: Payment::"P0"
  context: {
    isBusinessHours: false,
    onCorporateNetwork: true
  }
  entities: [
    Person::"P0" {
      isFinanceManager: true,
      ...
    },
    Payment::"P0" {
      invoice: Invoice::"I0",
      ...
    }
  ]
\end{raiseprompt}

\begin{raiseprompt}
{RaiseUnder}
{Over-restrictive policy: on-network floor violation}
{lst:raise-overrestrictive-network}
A verifier found your earlier attempts violated this requirement;
make sure your policy satisfies it:
- A finance manager must be allowed to issue a payment from the
  corporate network without being blocked merely because the
  invoice is not approved.
  concrete case your earlier attempt decided wrongly:
  principal: Person::"P0"
  action: Action::"issuePayment"
  resource: Payment::"P0"
  context: {
    isBusinessHours: true,
    onCorporateNetwork: true
  }
  entities: [
    Person::"P0" {
      isFinanceManager: true,
      ...
    },
    Payment::"P0" {
      invoice: Invoice::"I0",
      ...
    },
    Invoice::"I0" {
      status: "",
      ...
    }
  ]
\end{raiseprompt}

\begin{raiseprompt}
{RaiseLive}
{Deny-all policy: liveness violation}
{lst:raise-liveness}
A verifier found your earlier attempts violated this requirement;
make sure your policy satisfies it:
- At least one finance-manager issue-payment request that satisfies
  the visible rules should remain possible.
  concrete case your earlier attempt decided wrongly:
  LIVENESS VIOLATION: candidate policy is disjoint from the
  required liveness probe, so no approved request slice remains
  possible.
\end{raiseprompt}

\newpage

\newpage
\section{Construction of CedarInstruct}
\label{sec:dataset-construction}

We construct CedarInstruct, a dataset \(\mathcal{E}=\{e_i\}_{i=1}^{5800}\) of verified access control scenarios spanning \(19\) application areas and \(44\) domains. Each scenario is generated from a sampled structured specification \(\omega_i\) and stored as a tuple \(e_i=(D_i,\Sigma_i,\Pi_i,P_i^\star,\mathcal{R}_i)\), where \(D_i\) is the natural-language requirement, \(\Sigma_i\) is the Cedar schema, \(\Pi_i\) is the symbolic boundary plan, \(P_i^\star\) is the target Cedar policy bundle, and \(\mathcal{R}_i\) is the set of reference-policy files used by the executable checks.

\begin{figure}[t]
    \centering
    \includegraphics[width=\textwidth]{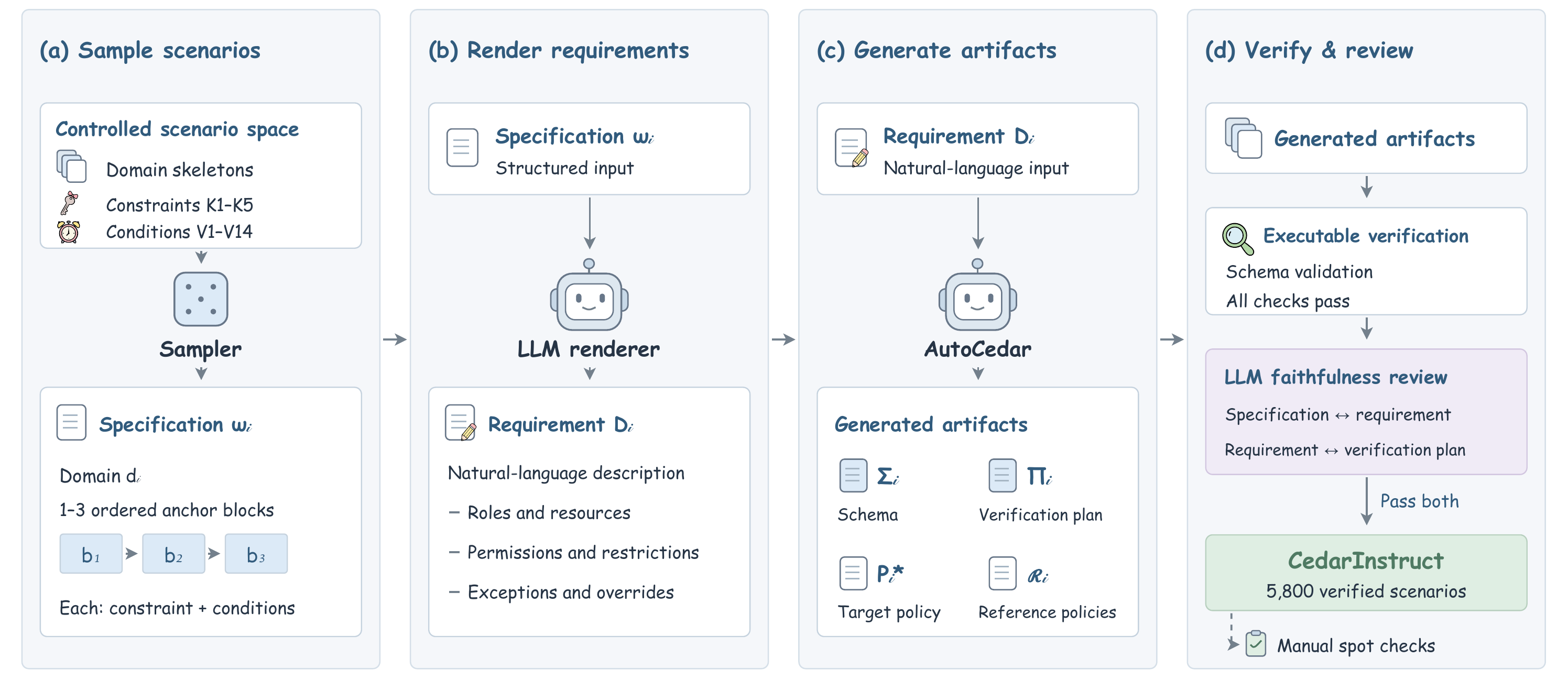}
    \caption{Four-stage construction pipeline of CedarInstruct.}
    \label{fig:dataset-pipeline}
\end{figure}

We build each scenario \(e_i\) through the four-stage pipeline shown in Figure~\ref{fig:dataset-pipeline}. First, the sampler draws a structured specification \(\omega_i\) from a domain-constrained scenario space. Second, an LLM renders \(\omega_i\) into a natural-language requirement \(D_i\). Third, AutoCedar uses \(D_i\) to generate the Cedar schema \(\Sigma_i\), boundary plan \(\Pi_i\), target policy bundle \(P_i^\star\), and reference policies \(\mathcal{R}_i\). Finally, every generated scenario undergoes executable validation. To assess construction faithfulness, we additionally conduct an LLM-based review on a sample of the retained scenarios and manually audit a subset of the LLM-reviewed sample using the same criteria. Thus, every retained scenario passes executable verification, whereas faithfulness is assessed through sampled LLM and human review rather than exhaustive per-scenario review.

\subsection{Stage 1: Controlled Scenario Space and Sampling}
\label{sec:scenario-space}

A structured specification \(\omega_i\) is the machine-readable source of a scenario. We represent it as
\[
\omega_i=(d_i,\mathbf{b}_i), \qquad \mathbf{b}_i=(b_{i,1},\ldots,b_{i,s_i}),
\]
where \(d_i\) identifies a domain skeleton and \(\mathbf{b}_i\) is an ordered sequence of anchor rule blocks. The domain skeleton provides the declared set of actions supported by the domain. Each block \(b_{i,j}\) selects one of these actions as an anchor and specifies its structural constraint, attached conditions, and any adjudication needed to resolve an override. Only the selected anchor actions are governed by the scenario; the remaining supported actions provide domain context rather than additional access-control requirements. This representation fixes the governed actions and their rule structure before any prose is generated, while the domain skeleton ensures that all referenced entities and fields can be represented in the Cedar schema \(\Sigma_i\).

Each domain is represented by a hand-written skeleton. The skeleton defines the roles, resources, typed fields, request-context fields, supported actions, and admissible rule structures. It also records which actions can form a linked workflow over the same resource. In such workflows, later actions read state stored as resource fields maintained by the application. This keeps multi-action scenarios compatible with Cedar's stateless request semantics.

\begin{table}[t]
\centering
\small
\setlength{\tabcolsep}{5pt}
\renewcommand{\arraystretch}{1.12}
\caption{Constraint categories used by the scenario sampler.}
\label{tab:constraint-categories}
\begin{tabular*}{\columnwidth}{@{\extracolsep{\fill}}l p{3cm} p{6.8cm}@{}}
\toprule
\textbf{ID} & \textbf{Constraint category} & \textbf{Semantic relation} \\
\midrule
K1 & Separation of duty & The current actor must differ from the actor stored for a conflicting prior action. \\
K2 & Binding of duty & The current actor must match the actor stored for a related prior action. \\
K3 & Override & An exception case takes precedence over the default access-control rule. \\
K4 & Level dominance & The current actor's authority level must dominate the resource's sensitivity level. \\
K5 & Lifecycle-state gate & The current action is allowed only in selected resource lifecycle states. \\
\bottomrule
\end{tabular*}
\end{table}

\begin{table}[t]
\centering
\small
\setlength{\tabcolsep}{5pt}
\renewcommand{\arraystretch}{1.08}
\caption{Condition types used by the scenario sampler.}
\label{tab:condition-types}
\begin{tabular*}{\columnwidth}{@{\extracolsep{\fill}}l p{3.1cm} p{6cm}@{}}
\toprule
\textbf{ID} & \textbf{Condition type} & \textbf{Plain meaning} \\
\midrule
\multicolumn{3}{@{}l}{\textit{\textbf{Identity and relationships}}} \\
V1 & Entity equality & The actor matches an identity field. \\
V2 & Type or role test & The actor has a required role or kind. \\
V3 & Hierarchy membership & The actor belongs to a group or hierarchy. \\
V4 & Set containment & The actor appears in an allowed set. \\
\addlinespace[2pt]
\multicolumn{3}{@{}l}{\textit{\textbf{Values and comparisons}}} \\
V5 & String condition & A text field matches a value or prefix. \\
V6 & Integer comparison & An integer field crosses a threshold. \\
V7 & Arithmetic condition & A computed value satisfies a threshold. \\
V8 & Decimal comparison & A precise numeric value satisfies a threshold. \\
V9 & Network condition & The request comes from an allowed network. \\
V10 & Time condition & The request satisfies a time or expiry condition. \\
\addlinespace[2pt]
\multicolumn{3}{@{}l}{\textit{\textbf{State, presence, and context}}} \\
V11 & Enum status & A resource is in a required lifecycle state. \\
V12 & Optional field & An optional field is present or absent. \\
V13 & Tags & A resource carries a required label. \\
V14 & Request context & The request environment satisfies a condition. \\
\bottomrule
\end{tabular*}
\end{table}

The sampler builds each anchor block from two controlled vocabularies. Constraint categories K1--K5 describe the structural access-control constraints assigned to an anchor action (Table~\ref{tab:constraint-categories}). Condition types V1--V14 describe additional predicates attached to those constraints (Table~\ref{tab:condition-types}). We include constraints evaluable from the current request and stored entity fields. Requirements involving prior actions are included only when the relevant state is already exposed by the domain skeleton. We exclude requirements that need access to unrepresented execution history, global aggregation, or reasoning about state transitions. This restriction does not exclude a separation-of-duty check that compares the current actor with a stored prior actor.

We describe each specification by its scope, depth, and form. The scope is the number of anchor blocks, \(s_i=|\mathbf{b}_i|\). A scope-one scenario contains one anchor action, while scope-two and scope-three scenarios contain linked action sequences declared by the domain skeleton. The depth \(\ell_i\) counts rule elements across all anchor blocks, including their structural constraints and attached conditions. The form \(f_i\) describes how these elements combine. Scenarios with an explicit override are classified as override-based; among the remainder, those with at least one carve-out are exception-based, and those with restriction-only conditions are conjunctive. These labels describe scenario structure rather than measured model difficulty.

The sampler constructs \(\omega_i\) directly in this space. It first selects a domain \(d_i\), then chooses \(\mathbf{b}_i\) as either a single anchor block or a linked sequence of two or three blocks supported by the domain skeleton. For each block, it selects a constraint category and attaches condition types whose required fields exist in the skeleton and whose combinations are valid under Cedar typing. Multi-action scenarios can only use workflow state exposed by the skeleton as resource fields. Before natural-language rendering, the sampler rejects any specification whose structured content duplicates a previously retained scenario. The retained scenarios therefore remain grounded in their domains while covering distinct combinations of actions, constraints, and conditions.

The condition types describe the meaning of the sampled predicates, not a one-to-one classification of Cedar operators. For example, a time or network condition can be represented by a Boolean context field supplied by the application. Coverage of V9 and V10 therefore does not by itself establish coverage of Cedar's IP-address or datetime operations.

\subsection{Stage 2: Natural-Language Requirement Rendering}
\label{sec:nl-rendering}

Given a sampled specification \(\omega_i\), we use GPT-5.5 as a renderer to produce the natural-language requirement \(D_i\). The renderer is called with a fixed prompt \(p_{\mathrm{render}}\) and the sampled specification, written \(D_i=\texttt{gpt5.5}(p_{\mathrm{render}},\omega_i)\). The prompt asks the model to draft an access-control requirement for a realistic organization while preserving the selected anchor actions, the roles, entities, and fields referenced by their rules, and the sampled constraint and condition relationships. Actions listed only as part of the domain inventory provide context and need not be restated as requirements. The prompt also defines the input format, output structure, vocabulary restrictions, conflict-resolution requirements, a missing-field rejection rule, and few-shot examples.

\subsection{Stage 3: Cedar Artifact Generation}
\label{sec:artifact-generation}

Given the rendered requirement \(D_i\), AutoCedar~\citep{vatsa2026autocedar} generates a Cedar schema \(\Sigma_i\) containing the entity types, fields, actions, and request context needed to express the requirement. It decomposes \(D_i\) into atomic access-control properties and uses an LLM to assign each property a verification relation. The resulting boundary plan \(\Pi_i=(\mathcal{C}_i,\mathcal{F}_i,\mathcal{G}_i)\) is stored as executable checks with natural-language descriptions and declared request scopes. The reference policies used by these checks are stored in \(\mathcal{R}_i\).

The checks implement the relations defined in Equation~\ref{eq:check-conditions}. A ceiling or floor reference policy \(R\) specifies an upper or lower bound through its permitted request set \(\operatorname{sem}_{\Sigma_i}(R)\). For a reference-based liveness check, the corresponding slice is \(G=\operatorname{sem}_{\Sigma_i}(R)\), and the candidate must authorize at least one request in this slice. Each check is evaluated within its declared principal, action, and resource scope and the input space supported by the symbolic encoding. Request evaluation includes the entity data needed to interpret policy conditions.

The property-atomization prompt assigns paired floor and ceiling checks to exact authorization conditions. Conditions stated only as sufficient produce floor checks, while conditions stated only as necessary produce ceiling checks. Within a shared action and request scope, every floor region must lie within every applicable ceiling region. AutoCedar then searches for \(P_i^\star\) through verifier-guided generation and repair, keeping the admitted schema and boundary plan fixed during candidate search.

\subsection{Stage 4: Verification and Faithfulness Filtering}
\label{sec:verification-filtering}

Stage~4 combines a dataset-wide executable verification filter with a sampled faithfulness audit. For executable verification, we validate \(P_i^\star\) and every reference policy in \(\mathcal{R}_i\) under \(\Sigma_i\), and execute all checks in \(\Pi_i\) using \texttt{cedar symcc}. A scenario passes this filter only when validation succeeds and \(P_i^\star\models_{\Sigma_i}\Pi_i\).

After executable filtering, we sample retained scenarios for LLM-based faithfulness review. Given \((\omega_i,D_i,\Pi_i)\), the judge assesses whether \(D_i\) preserves the sampled anchor rules in \(\omega_i\) without introducing unsupported requirements, and whether the properties described in \(\Pi_i\) are supported by and sufficiently cover the enforceable requirements in \(D_i\). Liveness checks are assessed for consistency with \(D_i\), rather than requiring a corresponding sentence in the requirement.

We manually review a subset of the LLM-reviewed scenarios using the same criteria. The manual audit compares the structured specification with the rendered requirement and compares the requirement with the generated boundary plan and its reference policies. Symbolic verification establishes satisfaction of the generated plan within the modeled input space; agreement with the natural-language requirement is assessed through sampled faithfulness review rather than proved by the solver.

\subsection{Dataset Characterization}
\label{sec:dataset-characterization}

\begin{figure}[htbp]
    \centering
    \includegraphics[width=0.7\textwidth]{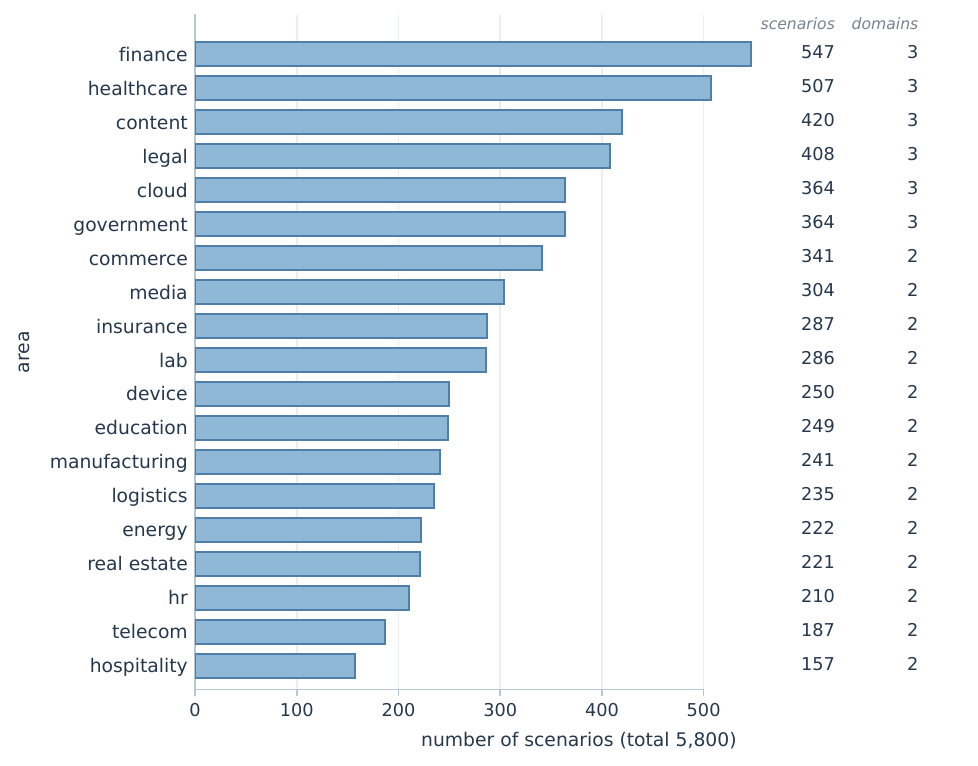}
    \caption{Number of retained scenarios in each application area. Domain and anchor-action coverage are reported in the text.}
    \label{fig:area-coverage}
\end{figure}

\paragraph{Coverage at multiple granularities.}
Figure~\ref{fig:area-coverage} shows the area-level distribution of the retained dataset. The \(5{,}800\) scenarios span all \(19\) application areas, with area sizes ranging from \(157\) to \(547\) scenarios. At the domain level, all \(44\) domain skeletons are covered, with \(64\) to \(233\) scenarios per domain and a median of \(130\). Every anchor action declared by each domain skeleton appears in the retained dataset, covering all \(108\) supported \((\mathrm{domain},\mathrm{anchor})\) pairs. These counts describe coverage of the declared skeletons, not of access-control applications in general.

\paragraph{Scenario structure.}
The dataset is concentrated in single-action scenarios, with \(4{,}974\) scope-one scenarios, \(794\) scope-two scenarios, and \(32\) scope-three scenarios. Depth ranges from \(1\) to \(9\), with \(76.1\%\) of scenarios at depth \(2\) or \(3\). Under the mutually exclusive form classification, \(3{,}093\) scenarios are conjunctive, \(2{,}022\) are exception-based, and \(685\) are override-based. All five constraint categories and all fourteen condition types appear in the retained dataset. The condition types appear \(9{,}809\) times as attached conditions, with \(69.5\%\) used as restrictions and \(30.5\%\) used as exceptions.

\paragraph{Verification density.}
Across the dataset, the boundary plans contain \(44{,}388\) checks, with a mean of \(7.65\), a median of \(7\), and a 90th percentile of \(12\) checks per scenario. Every scenario contains at least one ceiling and one floor check. Ceiling checks account for \(23{,}909\) checks (\(53.9\%\)), floor checks account for \(17{,}331\) checks (\(39.0\%\)), and liveness checks account for the remaining \(3{,}148\) checks (\(7.1\%\)). These checks measure over-permission and over-restriction relative to the generated boundaries, without requiring textual similarity to \(P_i^\star\). Check count describes the size of the verification plan, not necessarily its semantic coverage or difficulty.

\paragraph{Splits.} We assign each scenario to exactly one partition using a fixed random seed while preserving coverage across application areas and domains. This produces \(4{,}386\) SFT examples, \(1{,}039\) RL examples, and \(375\) test examples. The assignment is fixed before model training and is never changed during SFT-data selection, RL-data selection, model selection, or evaluation. The test split is used only for final in-distribution evaluation and is not used for training, hyperparameter tuning, or checkpoint selection. It covers all \(19\) application areas and all \(44\) domains, with \(11\) to \(35\) scenarios per area and \(5\) to \(14\) scenarios per domain. Because splitting is performed at the scenario level, this is not a schema-, action-, or domain-level out-of-distribution split.

\subsection{Example Retained Scenario}
\label{sec:dataset-example}

We illustrate the construction pipeline using scenario \texttt{invoice\_approval-issuePayment-T3-045} from the invoice-approval domain. It has scope \(1\), depth \(3\), and combines a lifecycle-state gate (K5), a business-hours restriction (V10), and a network exception (V9) on \texttt{issuePayment}. For this presentation, the specification and requirement have been clarified to state the allowed and denied cases explicitly, and the boundary plan is shown as a readable checklist. The schema, target policy, and reference policy bodies are unchanged. Each reference policy is stored in a separate file.

\begin{datasetblock}{Scenario specification}{lst:example-spec}
DOMAIN:
     finance / invoice approval

ROLES:
     clerks, approvers, finance managers, auditors; a single person may hold both the clerk and approver roles.

RESOURCES:
     invoice records its creator, approver, amount, surcharge, line item count, status, cost center, submission time; payment records its invoice, issuer, amount.

REQUEST CONTEXT:
     each request records whether it is made during business hours, whether it originates from the corporate network, and whether an emergency has been declared.

ACTIONS:
     create invoice, approve invoice, issue payment, view audit log.

ANCHOR ACTION:
     issue a payment (finance managers only)

RULES (all bear on the anchor action):
- Subject to the restriction and exception below, finance managers must be allowed to issue 
payments for approved invoices and must be denied for unapproved invoices. [K5]
- only-when: payments may be issued only during business hours; outside business hours, every 
payment request must be denied. [V10]
- unless: on the corporate network, the approved-invoice requirement is lifted. During business 
hours, finance managers must be allowed to issue payments from that network regardless of invoice 
approval. This exception does not waive the role or business-hours requirement. [V9]
\end{datasetblock}

\begin{datasetblock}{Natural-language requirement}{lst:example-requirement}
# Invoice Payment

## Natural-Language Scenario

The system manages and stores invoice approval and payment records.

The people involved are clerks, approvers, finance managers, and auditors. 

One person may hold both the clerk and approver roles at the same time.

Every invoice records its creator, approver, amount, surcharge, line item count, status, cost 
center, and submission time.

Every payment records its invoice, issuer, and amount.

Each payment request records whether it is made during business hours, whether it originates from 
the corporate network, and whether an emergency has been declared.

Anyone who is not a finance manager must be denied permission to issue a payment.

Outside business hours, every request to issue a payment must be denied.

During business hours and outside the corporate network, a finance manager must be allowed to issue
a payment when the invoice has been approved; the manager must be denied permission to issue a 
payment when the invoice has not been approved. 

During business hours and on the corporate network, a finance manager must be allowed to issue a 
payment regardless of whether the invoice has been approved.
\end{datasetblock}

\begin{datasetblock}{Cedar schema}{lst:example-schema}
entity Person { isClerk: Bool, isApprover: Bool, isFinanceManager: Bool, isAuditor: Bool };

entity Invoice { creator: Person, approver: Person, amount: decimal, surcharge: decimal,
                 lineItemCount: Long, status: String, costCenter: String,
                 submissionTime: datetime };

entity Payment { invoice: Invoice, issuer: Person, amount: decimal };

action issuePayment appliesTo {
    principal: [Person],
    resource: [Payment],
    context: { onCorporateNetwork: Bool, emergencyDeclared: Bool, isBusinessHours: Bool },
};
\end{datasetblock}

\begin{datasetblock}{Target policy bundle}{lst:example-policy}
permit (principal is Person, action == Action::"issuePayment", resource is Payment)
when {
    principal.isFinanceManager
    && context.isBusinessHours
    && (context.onCorporateNetwork || resource.invoice.status == "approved")
};
\end{datasetblock}

\begin{datasetblock}{Boundary plan}{lst:example-boundary-plan}
REQUEST SHAPE
Principal: Person
Action: Action::"issuePayment"
Resource: Payment

[F1] FLOOR
Name: finance_manager_issue_approved_invoice_off_network_floor
Description: During business hours, a finance manager must be allowed to issue a payment when the 
request is outside the corporate network and the invoice has been approved.
Reference: finance_manager_issue_approved_invoice_off_network_floor.cedar

[F2] FLOOR
Name: finance_manager_issue_payment_on_network_floor
Description: During business hours, a finance manager must be allowed to issue a payment from the 
corporate network regardless of invoice approval.
Reference: finance_manager_issue_payment_on_network_floor.cedar

[C1] CEILING
Name: finance_manager_issue_payment_network_status_ceiling
Description: A payment may be issued only by a finance manager. Outside the corporate network, the 
invoice must also be approved. On the corporate network, invoice approval is not required by this 
boundary.
Reference: finance_manager_issue_payment_network_status_ceiling.cedar

[C2] CEILING
Name: finance_manager_issue_payment_business_hours_disjointness
Description: No request to issue a payment may be allowed outside business hours.
Reference: finance_manager_issue_payment_business_hours_disjointness.cedar

[L1] LIVENESS
Name: finance_manager_issue_payment_clean_request_liveness
Description: At least one request by a finance manager during business hours must be allowed, with 
either a corporate-network origin or an approved invoice.
Reference: finance_manager_issue_payment_clean_request_liveness.cedar
\end{datasetblock}

\begin{datasetblock}{Check-specific reference policies}{lst:example-reference-policies}
// [F1] finance_manager_issue_approved_invoice_off_network_floor.cedar

permit (principal, action == Action::"issuePayment", resource) when {
    (principal is Person && resource is Payment && principal.isFinanceManager
     && !context.onCorporateNetwork && resource.invoice.status == "approved")
    && !(!context.isBusinessHours)
};

// [F2] finance_manager_issue_payment_on_network_floor.cedar

permit (principal, action == Action::"issuePayment", resource) when {
    (principal is Person && resource is Payment && principal.isFinanceManager
     && context.onCorporateNetwork)
    && !(!context.isBusinessHours)
};

// [C1] finance_manager_issue_payment_network_status_ceiling.cedar

permit (principal, action == Action::"issuePayment", resource) when {
    principal is Person && resource is Payment && principal.isFinanceManager
    && (context.onCorporateNetwork || resource.invoice.status == "approved")
};

// [C2] finance_manager_issue_payment_business_hours_disjointness.cedar

permit (principal is Person, action == Action::"issuePayment", resource is Payment)
when {
    !(!context.isBusinessHours)
};

// [L1] finance_manager_issue_payment_clean_request_liveness.cedar

permit (principal, action == Action::"issuePayment", resource) when {
    principal is Person && resource is Payment && principal.isFinanceManager
    && context.isBusinessHours
    && (context.onCorporateNetwork || resource.invoice.status == "approved")
};
\end{datasetblock}
\subsection{Reading a Cedar Policy}
\label{app:cedar-example}

A Cedar policy bundle is a set of \texttt{permit} and \texttt{forbid}
rules. Each rule names the principals, actions, and resources it applies
to, and may add conditions in \texttt{when} clauses (the rule applies
only if the condition holds) and \texttt{unless} clauses (the rule
applies only if it does not). A request is allowed if and only if at
least one \texttt{permit} rule applies and no \texttt{forbid} rule
applies. We illustrate with the invoice-payment scenario of
Appendix~\ref{sec:dataset-example}, which allows finance managers to
issue payments only during business hours, and only for approved
invoices unless the request comes from the corporate network.

The bundle below looks natural but is wrong. It writes the network
exception as a separate rule, and that rule omits the business-hours
condition, so a finance manager on the corporate network can issue a
payment at any time.

\begin{datasetblock}{A plausible but incorrect policy bundle}{lst:cedar-wrong}
permit (principal is Person, action == Action::"issuePayment", resource is Payment)
when {
  principal.isFinanceManager
  && context.isBusinessHours
  && resource.invoice.status == "approved"
};

permit (principal is Person, action == Action::"issuePayment", resource is Payment)
when {
  principal.isFinanceManager
  && context.onCorporateNetwork
};
\end{datasetblock}

The verifier rejects this bundle because it violates the business-hours
ceiling check C2 (Listing~\ref{lst:example-boundary-plan}); any
counterexample is a request by a finance manager on the corporate
network outside business hours. One correct fix keeps both
\texttt{permit} rules and adds a \texttt{forbid} rule that blocks every
payment outside business hours. The target policy in
Listing~\ref{lst:example-policy} is an equivalent single-rule version.

\begin{datasetblock}{A corrected policy bundle}{lst:cedar-fixed}
permit (principal is Person, action == Action::"issuePayment", resource is Payment)
when {
  principal.isFinanceManager
  && context.isBusinessHours
  && resource.invoice.status == "approved"
};

permit (principal is Person, action == Action::"issuePayment", resource is Payment)
when {
  principal.isFinanceManager
  && context.onCorporateNetwork
};

forbid (principal, action == Action::"issuePayment", resource)
unless { context.isBusinessHours };
\end{datasetblock}

\newpage

\section{Implementation Hyperparameters}
\label{sec:implementation-details}

\begin{table}[htbp]
\centering
\cedartablestyle
\fontsize{8.2}{8.9}\selectfont
\setlength{\tabcolsep}{4pt}
\setlength{\extrarowheight}{0pt}
\renewcommand{\arraystretch}{0.96}
\caption{Training hyperparameters for the SFT and \textsc{RAISE} runs using the \texttt{verl} reproduction framework. Qwen3.5-9B SFT was run for four epochs to assess saturation, while Qwen3.8-27B SFT was run for three epochs. All reported SFT results and subsequent \textsc{RAISE} runs use the final epoch-3 SFT checkpoint.}
\label{tab:training-hyperparameters}
\begin{tabularx}{\linewidth}{@{}Xcc@{}}
\toprule
\rowcolor{cedarHeader}
\textbf{Hyperparameter} & \textbf{Qwen3.5-9B} & \textbf{Qwen3.8-27B} \\
\midrule
\multicolumn{3}{c}{\textit{Supervised fine-tuning}} \\
\midrule
Adaptation & LoRA & LoRA \\
LoRA rank $r$ & 64 & 64 \\
LoRA scaling $\alpha$ & 128 & 128 \\
LoRA dropout & 0.05 & 0.05 \\
SFT epochs run & 4 & 3 \\
Optimizer updates & 414 & 414 \\
Optimizer & Fused AdamW & Fused AdamW \\
Peak learning rate & $10^{-4}$ & $10^{-4}$ \\
Scheduler & Cosine & Cosine \\
Warmup ratio & 0.10 & 0.10 \\
Weight decay & 0.01 & 0.01 \\
Global batch size & 32 & 32 \\
Examples per GPU & 2 & 2 \\
Gradient accumulation & 4 & 4 \\
Maximum sequence length & 8{,}192 & 8{,}192 \\
Checkpoint used for reporting and \textsc{RAISE} initialization & Epoch 3 & Epoch 3 \\
\midrule
\multicolumn{3}{c}{\textit{\textsc{RAISE} reinforcement learning}} \\
\midrule
Adaptation & LoRA & LoRA \\
LoRA rank $r$ & 32 & 32 \\
LoRA scaling $\alpha$ & 64 & 64 \\
LoRA dropout & 0.05 & 0.05 \\
RL epochs & 2 & 2 \\
Optimizer updates & 518 & 518 \\
Optimizer & Fused AdamW & Fused AdamW \\
Learning rate & $10^{-5}$ & $10^{-5}$ \\
Scheduler & Constant & Constant \\
Warmup steps & 0 & 0 \\
Weight decay & 0 & 0 \\
Global prompt batch size & 4 & 4 \\
Initial rollouts $K$ & 8 & 8 \\
Additional guided rollouts & 8 & 8 \\
Feedback limit $M$ & 4 & 4 \\
Sampling temperature & 0.9 & 0.9 \\
Top-$p$ & 1.0 & 1.0 \\
Top-$k$ filtering & Disabled & Disabled \\
Maximum completion length & 3{,}072 & 3{,}072 \\
Optimization iterations & 1 & 1 \\
Clipping parameter $\epsilon$ & 0.2 & 0.2 \\
Advantage standard deviation & Sample & Sample \\
Advantage stabilizer $\delta$ & $10^{-4}$ & $10^{-4}$ \\
KL coefficient $\beta$ & 0.01 & 0.01 \\
KL reference & Fixed SFT policy & Fixed SFT policy \\
Training seed & 42 & 42 \\
\bottomrule
\end{tabularx}
\end{table}

\newpage

\section{Baseline Methods}
\label{sec:baseline-methods}

All trainable baselines use the CedarInstruct RL partition and start from the same backbone-specific SFT checkpoint as \textsc{RAISE}. Their common input is $q=(D,\Sigma)$, consisting of the natural-language requirement and Cedar schema. Reward construction, rollout composition, optimization objectives, and update budgets vary by method, as detailed below and in Appendix~\ref{sec:implementation-details}. Evaluation uses greedy single-shot generation conditioned only on $q$, without verifier feedback or iterative repair.

\paragraph{Prompting Baselines.}
Zero-shot prompting directly asks the pretrained backbone to generate a Cedar policy from the requirement and schema. Structured zero-shot prompting additionally specifies the expected output format and Cedar-generation constraints, without parameter updates or verifier-assisted repair. The prompts are provided in Appendix~\ref{app:prompts}. GPT-6 Astra (Max)~\citep{openai2026gpt6astra} and Claude Opus 5 (Max)~\citep{anthropic2026claudeopus5} are evaluated as independent zero-shot frontier references. GPT-5.5 (Medium)~\citep{openai2026gpt55}, which serves as the LLM backend for model-mediated stages of CedarInstruct construction, is reported separately as a construction-model reference.

\paragraph{Binary Reward GRPO.}
Binary Reward GRPO applies group-relative policy optimization~\citep{shao2024deepseekmath} using the same full-verification outcome as \textsc{RAISE}. Let $\operatorname{Valid}_{\Sigma}(P)$ indicate that $P$ parses and passes schema validation. The reward is
\[
r_{\mathrm{bin}}(P)
=
\mathbb{I}
\!\left[
\operatorname{Valid}_{\Sigma}(P)
\land
P\models_{\Sigma}\Pi
\right].
\]
Candidates are sampled under the original input, and rewards are normalized within each prompt group to obtain relative advantages. The method performs no feedback-guided regeneration. Consequently, all-failure and all-success groups have zero outcome-based advantages.

\paragraph{Coverage Reward GRPO.}
Coverage Reward GRPO tests whether a denser scalar reward can provide useful supervision when no candidate satisfies the complete plan. For a schema-valid candidate, let $\operatorname{Pass}(P)\subseteq\Pi$ be its satisfied checks and $\phi(P)=|\operatorname{Pass}(P)|/|\Pi|$. We mark a partial candidate as degenerate when it passes no ceiling check, or passes neither a floor nor a liveness check:
\[
d(P)
=
\mathbb{I}
\!\left[
\bigl(\mathcal C\neq\varnothing
\land
\operatorname{Pass}(P)\cap\mathcal C=\varnothing\bigr)
\lor
\bigl(\mathcal F\cup\mathcal G\neq\varnothing
\land
\operatorname{Pass}(P)\cap(\mathcal F\cup\mathcal G)=\varnothing\bigr)
\right].
\]
The guarded coverage reward is
\[
r_{\mathrm{cov}}(P)
=
\begin{cases}
0,
& P\text{ does not parse},\\
0.10,
& P\text{ parses but fails schema validation},\\
1,
& P\models_{\Sigma}\Pi,\\
0.15,
& d(P)=1,\\
0.15+0.65\,\phi(P),
& \text{otherwise}.
\end{cases}
\]
The guard penalizes verification patterns associated with excessively permissive or restrictive policies; it does not establish that a candidate literally permits or denies every request. This baseline provides partial credit but does not use failed-check descriptions or counterexamples for further generation.

\paragraph{Formal Dominance GRPO.}
Formal Dominance GRPO extracts pairwise preferences from the identities of satisfied checks. Candidate $P_a$ strictly dominates $P_b$ when
$\operatorname{Pass}(P_b)\subsetneq\operatorname{Pass}(P_a)$; candidates with incomparable passed-check sets receive no preference relation. Mixed-reward groups use the binary GRPO objective. For an all-failure group, syntax-invalid candidates are removed and valid candidates with identical passed-check sets form equivalence classes. We retain only covering dominance relations, removing relations implied transitively through another class.

Let $\mathcal H_q$ denote the retained class relations and let $s_\theta(P\mid q)$ be the policy sequence score under the original input. The dominance loss is
\[
\mathcal L_{\mathrm{dom}}
=
-\frac{1}{|\mathcal H_q|}
\sum_{(A,B)\in\mathcal H_q}
\frac{1}{|A||B|}
\sum_{\substack{P_a\in A\\P_b\in B}}
\log\sigma
\!\left(
s_\theta(P_a\mid q)-s_\theta(P_b\mid q)
\right).
\]
The ranking scale and dominance-loss weight are both $1.0$. Groups without a strict dominance relation provide no ranking signal, and all-success groups are skipped. The reference-policy KL regularizer is applied only to the mixed-group GRPO branch. This baseline tests whether preferences among existing failed candidates suffice without feedback-conditioned exploration.

\paragraph{Critique-GRPO.}
We implement a Cedar-specific adaptation of Critique-GRPO~\citep{zhang2025critique}. Each prompt produces six initial candidates, and failed candidates with verifier feedback undergo up to two refinement rounds. Each round uses at most two failed-check descriptions, prioritizing over-permissive before over-restrictive failures; no separate critic or counterexample is used. We retain two refinements, prioritizing correct candidates, and combine them with the six initial candidates. If fewer than two refinements are available, the best available refinement or an initial candidate is reused. Binary rewards are centered over the combined group without standard-deviation normalization. Initial candidates use the clipped policy-gradient objective, while refinement tokens use the shaping factor $p/(p+0.1)$ under the original input, without context correction or reference-policy KL regularization.

\paragraph{SDPO.}
We adapt SDPO~\citep{hubotter2026reinforcement} by conditioning the teacher on the original input, an available successful sibling policy, and verifier descriptions of the sampled candidate's failures; the student receives only the original input. Teacher and student score the same completion tokens, and the teacher adapter is updated from the student by exponential moving average with rate $0.05$. We minimize forward KL over the student's top $20$ tokens and a tail bucket, with token-level importance weights capped at $2.0$. This loss replaces GRPO and uses no reference-policy KL penalty. The 9B configuration requires both a successful sibling and verifier feedback, whereas the 27B configuration requires either one.

\section{Additional Experimental Results}

\subsection{Construction-Model Reference}
\label{app:construction-ref}

GPT-5.5 (Medium) served as the LLM backend for all model-mediated stages of CedarInstruct construction. For transparency, we report its zero-shot performance on the CedarInstruct test split; because it participated in constructing the dataset and its verification artifacts, we exclude it from comparisons. It achieves 82.93\% syntax, 38.13\% semantic, 66.22\% macro, and 63.60\% micro success, using on average 12,449 input and 606 output tokens per scenario, with 14.0 seconds latency and an estimated cost of \$0.0291.

\subsection{Metric Definitions}
\label{app:metrics}

Syntax is the percentage of policies that parse and pass schema validation. Semantic, our primary metric, is the percentage that are valid and satisfy every verification check. Macro averages the per-scenario fraction of passed checks, and Micro divides total passed checks by total checks. Invalid policies receive no semantic, macro, or micro credit.

\subsection{Significance Tests}
\label{app:significance}

On the 375-scenario test split, each comparison between two models is paired: both are evaluated on the same scenarios. We report, for RAISE-OC against SFT and against the strongest other instantiation on each backbone, a two-sided exact McNemar test on per-scenario semantic success and a 95\% paired bootstrap interval for the difference (10,000 resamples) These tests capture evaluation noise, not variance across training seeds.

\subsection{Guidance Ablations}
\label{app:guidance-ablation}

Table~\ref{tab:feedback-ablation} ablates RAISE-OC's guidance. Selecting checks by failure frequency outperforms random selection (46.93\% vs. 45.60\%); exploring each check in a separate guided context outperforms placing all selected feedback in one shared prompt (46.93\% vs. 44.80\%); and aggregating failures across candidates outperforms giving each rollout its own candidate's failures (46.93\% vs. 42.13\%), a gap that persists on Qwen3.8-27B (48.00\% vs. 45.87\%). The last comparison is the closest analogue in our experiments to guidance built from a single failed attempt \citep{agrawal2026off}, though both variants here use verifier feedback. Larger budgets did not help: $M=8$ with 16 guided rollouts reaches 45.60\%. Most configurations were run only on Qwen3.5-9B with a single seed, so we treat these comparisons as indicative.

\begin{table}[t]
\centering
\cedartablestyle
\fontsize{8.2}{9.0}\selectfont
\setlength{\tabcolsep}{5pt}
\renewcommand{\arraystretch}{0.95}
\caption{Guidance ablations for RAISE-OC: single-shot semantic success (\%). All feedback items contain a failed-check description and a counterexample; $M$ limits the number of selected checks. Every entry is a single greedy run on the 375-scenario test split.}
\label{tab:feedback-ablation}
\begin{tabularx}{\linewidth}{@{}c l X c cc@{}}
\toprule
\rowcolor{cedarHeader}
$M$ & \textbf{Check selection} & \textbf{Prompt layout} & \textbf{Rollouts} & \textbf{Qwen3.5-9B} & \textbf{Qwen3.8-27B} \\
\midrule
\rowcolor{cedarStripe}
1 & most frequent & one per check & 8 & 44.53 & 47.47 \\
2 & most frequent & one per check & 8 & 41.87 & 47.20 \\
\rowcolor{cedarStripe}
4 & most frequent & one per check & 8 & \textbf{46.93} & \textbf{48.00} \\
4 & random & one per check & 8 & 45.60 & 47.73 \\
\rowcolor{cedarStripe}
4 & most frequent & all in one shared prompt & 8 & 44.80 & 47.47 \\
8 & most frequent & one per check & 16 & 45.60 & 47.20 \\
\rowcolor{cedarStripe}
All & each rollout's failed checks & one per rollout & 8 & 42.13 & 45.87 \\
\bottomrule
\end{tabularx}
\end{table}

\subsection{Verifier Feedback at Inference Time}
\label{app:repair}

Table~\ref{tab:repair} applies verifier feedback at inference time. On every model, one round of repair with failed-check descriptions and counterexamples raises semantic success by 19 to 31 points, whereas a retry that reports only failure adds 2 to 7. The content of verifier feedback, not the extra attempt, drives most of the gain, which is the property RAISE-OC's guided exploration relies on during training. RAISE-OC also improves the starting point for repair: after two rounds it reaches 70.40\% on Qwen3.5-9B, versus 67.73\% for SFT. Trained models and verifier-guided repair are therefore complementary.

\begin{table}[t]
\centering
\begin{minipage}{\linewidth}
\cedartablestyle
\fontsize{8.2}{9.0}\selectfont
\caption{Feedback-assisted repair on the 375 test scenarios (\%). Repair supplies failed-check descriptions and available counterexamples; retry-only reports only that the policy failed.}
\label{tab:repair}
\begin{tabularx}{\linewidth}{Xcccc}
\toprule
\rowcolor{cedarHeader}
\textbf{Method} & \textbf{Initial} & \textbf{Repair 1} & \textbf{Repair 2} & \textbf{Retry only} \\
\midrule
\multicolumn{5}{c}{\textit{Qwen3.5-9B}} \\
\midrule
\rowcolor{cedarStripe}
SFT & 42.93 & 64.00 & 67.73 & 45.33 \\
Binary GRPO & 43.47 & 62.93 & 67.20 & 45.07 \\
\rowcolor{cedarStripe}
\textbf{RAISE-OC} & \textbf{46.93} & \textbf{66.93} & \textbf{70.40} & \textbf{50.40} \\
\midrule
\multicolumn{5}{c}{\textit{Qwen3.8-27B}} \\
\midrule
\rowcolor{cedarStripe}
SFT & 46.40 & 73.60 & 76.80 & 50.40 \\
Binary GRPO & 44.80 & \textbf{76.00} & \textbf{80.80} & 52.00 \\
\rowcolor{cedarStripe}
\textbf{RAISE-OC} & \textbf{48.00} & 74.40 & 78.93 & \textbf{52.27} \\
\bottomrule
\end{tabularx}
\end{minipage}
\end{table}

\begin{figure}[htbp]
\centering
\includegraphics[width=0.78\linewidth]{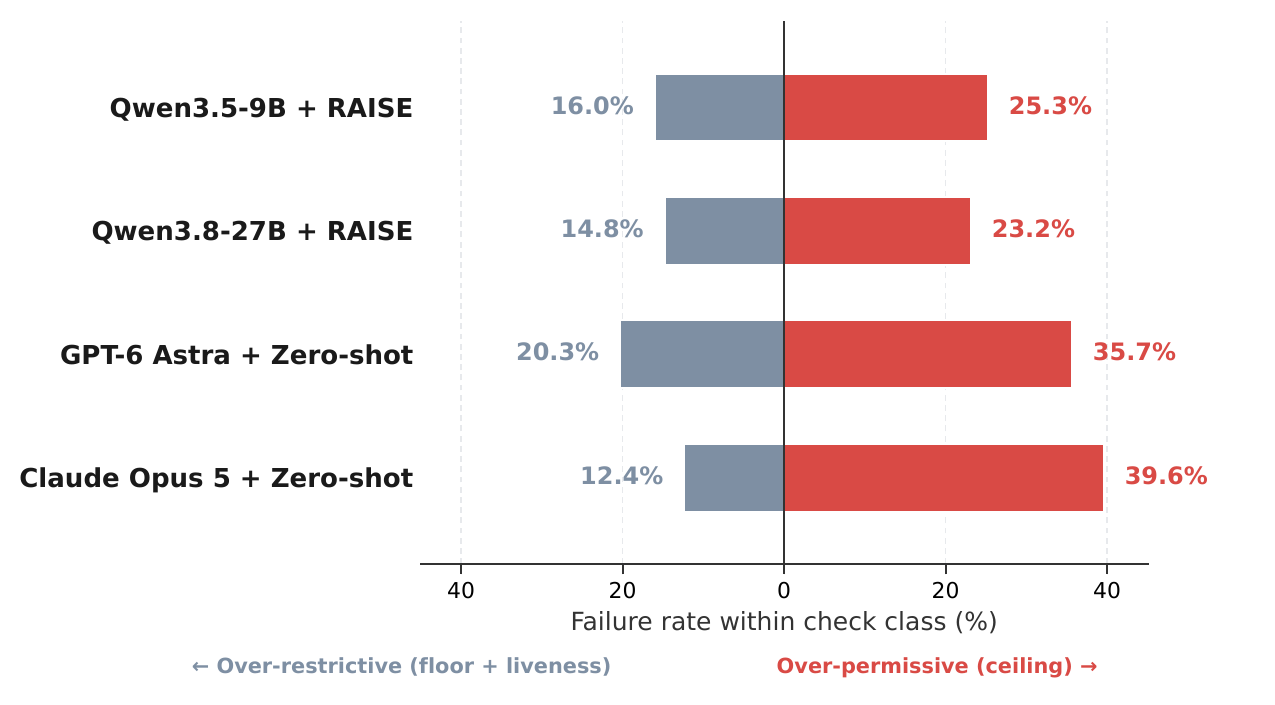}
\caption{Failure rates within over-restrictive and over-permissive check classes: failed checks divided by evaluated checks in each class.}
\label{fig:failure-patterns}
\end{figure}

\subsection{Authorization Failure Patterns}

Figure~\ref{fig:failure-patterns} reports failure rates within each check class. Ceiling checks fail more often than floor and liveness checks for every model shown, making over-permission the dominant failure mode. RAISE-OC has lower over-permissive failure rates (25.3\% and 23.2\%) than GPT-6 Astra (35.7\%) and Claude Opus 5 (39.6\%). Its over-restrictive rates (16.0\% and 14.8\%) fall below Astra's 20.3\% but above Opus's 12.4\%. Without SFT and Binary Reward GRPO for comparison, these rates do not attribute the reduction in over-permission to the RL stage specifically.

\subsection{Inference Efficiency}
\label{app:efficiency}

Table~\ref{tab:inference-efficiency} reports latencies of 3.85 and 7.35 seconds per scenario for the RAISE-OC models, versus 14.3 and 27.3 seconds for GPT-6 Astra and Claude Opus 5. The frontier figures are measured through each provider's agent CLI (Codex CLI and Claude Code), not as raw API calls: the task prompt itself is under 1K tokens, but an agent loop resends its full context on every call, so the reported input tokens are summed over all calls, include prompt-cache reads and writes, and contain a fixed harness context much larger than the prompt. Output tokens include reasoning tokens. At estimated costs of \$0.000093 and \$0.000370 per scenario, the RAISE-OC models are about $686\times$ and $172\times$ cheaper than GPT-6 Astra, the less expensive of the two frontier models in Table~\ref{tab:main}; because the frontier costs include this harness overhead, the ratios describe end-to-end deployment rather than bare API calls. Latency and cost depend on the provider, hardware, and prompt configuration, so these measurements characterize the evaluated deployments only.

\begin{table}[t]
\centering
\begin{minipage}{\linewidth}
\cedartablestyle
\fontsize{8.4}{8.9}\selectfont
\setlength{\tabcolsep}{2.5pt}
\setlength{\extrarowheight}{0pt}
\renewcommand{\arraystretch}{1}

\caption{Average per-scenario inference efficiency on the CedarInstruct test split. Lower is better except for Semantic; bold and underline mark the best and second-best results.}
\label{tab:inference-efficiency}
\vspace{-3pt}

\begin{tabularx}{\linewidth}{@{}l*{5}{>{\centering\arraybackslash}X}@{}}
\toprule
\rowcolor{cedarHeader}
\textbf{Model}
& \textbf{Input tokens}
& \textbf{Output tokens}
& \textbf{Latency (s)}
& \textbf{Cost (\$)}
& \textbf{Semantic (\%)} \\
\midrule
\rowcolor{cedarStripe}
GPT-6 Astra (Max)
& 24{,}832 & 448 & 14.3 & 0.0638 & 33.60 \\
Claude Opus 5 (Max)
& 49{,}134 & 2{,}412 & 27.3 & 0.1375 & 30.67 \\
\rowcolor{cedarStripe}
Qwen3.5-9B + \textsc{RAISE}
& \textbf{699}
& \underline{154}
& \textbf{3.85}
& \textbf{0.000093}\textsuperscript{$\dagger$}
& \underline{46.93} \\
Qwen3.8-27B + \textsc{RAISE}
& \underline{737}
& \textbf{141}
& \underline{7.35}
& \underline{0.000370}\textsuperscript{$\dagger$}
& \textbf{48.00} \\
\bottomrule
\end{tabularx}

\vspace{2pt}
{\fontsize{7}{7.8}\selectfont
$\dagger$ Qwen costs use DeepInfra rates listed by OpenRouter, not measured local deployment costs.\par}
\end{minipage}
\end{table}

\section{Limitations}
\label{app:limitations}

\paragraph{Correctness is relative to generated verification plans.} Symbolic checking establishes that a policy satisfies its verification plan, not that it matches every reading of the requirement. Plans are generated by AutoCedar with GPT-5.5, and their faithfulness to the requirements is assessed on a sample rather than exhaustively (Section~\ref{sec:cedarinstruct}). Because the same pipeline defines the RL reward, the feedback, and the test-set grading, trained models may partly learn the pipeline's reading of ambiguous requirements.

\paragraph{The test split is in-distribution at the domain level.} Test scenarios are distinct from training scenarios but share domain skeletons and constraint vocabularies. CedarBench tests transfer to independent scenarios, but we did not evaluate frontier models on it.

\paragraph{Statistical power.} All results come from a single training seed, and differences between RL instantiations are often a few scenarios. Paired tests on per-scenario outcomes (Appendix~\ref{app:significance}) address evaluation noise but not training variance.

\paragraph{Missing controls.} Guided exploration draws $K$ additional candidates for each all-failure prompt, so part of its benefit may come from extra samples; a resampling control, as in DAPO \citep{yu2026dapo}, would separate the two. We also did not compare verifier guidance with OC-GRPO's reference-solution prefixes or LLM-written hints; OC-GRPO's variance analysis, which favors short guidance, suggests verifier diagnostics should compare favorably, but we have not tested this. Finally, we did not run SFT on all 5,425 verified scenarios, which would show how much of the RL stage's benefit additional demonstrations could provide.

\paragraph{Dataset scope.} Most scenarios govern a single action (4,974 of 5,800). Time and network conditions are represented as Boolean context fields, so Cedar's datetime and IP-address operations are not exercised. Scenarios requiring execution history or global aggregation are excluded.

\paragraph{Answer-only supervision.} SFT targets contain only the policy. \citet{lin2025debunk} find that chain-of-thought supervision improves generalization to harder instances; the atomic properties in each verification plan offer a natural reasoning scaffold that we leave to future work.

\section{Prompt Templates}
\label{app:prompts}

Listing~\ref{lst:render-prompt} gives the rendering prompt used during dataset construction (Appendix~\ref{sec:dataset-construction}). Listings~\ref{lst:zero-shot-prompt} and~\ref{lst:structured-prompt} give the plain and structured zero-shot policy-generation prompts, respectively. The plain prompt in Listing~\ref{lst:zero-shot-prompt} is also used to construct both SFT and RL inputs: for SFT, the prompt is paired with the reference Cedar policy as the target response; for RL, the same prompt is used to elicit policy rollouts. Frontier models receive the zero-shot prompts.

\begin{datasetblock}{Rendering prompt}{lst:render-prompt}
Role:
You are a domain expert drafting an access-control requirements document for a real
organization. Write in plain, natural English. Do not write code or use technical access
-control terminology.

Input:
<structured specification>

Task:
Render <structured specification> as an ordered requirement document. Preserve every selected
anchor action and all roles, entities, fields, constraints, and condition relationships
used by its rules. The ACTIONS field lists the operations supported by the domain. Do
not turn non-anchor actions into additional requirements. For linked workflows, state
how later anchor actions depend on state stored by earlier steps. For exceptions or
overrides, state which cases follow the default rule and which cases are carved out or
take precedence.

Restrictions:
1. Use plain business English only.
2. Do not introduce roles, resources, fields, actions, or requirements outside <structured
   specification>.
3. If an action can be performed by several kinds of people, a prohibition must name all of
   them.

Fitness check:
If any rule needs a field not recorded by the domain, output exactly one line and stop:
REJECT: <missing field>

Examples:
<example_1>
<example_2>
<example_3>

Output:
# Title
## Natural-Language Scenario
<sentences>
\end{datasetblock}

\begin{datasetblock}{Policy-generation prompt used for zero-shot evaluation and SFT/RL training.}{lst:zero-shot-prompt}
SYSTEM MESSAGE
You are an expert Cedar access-control policy synthesizer. Given a Cedar schema and a natural
-language access-control requirement, write the Cedar policies that exactly implement it.

Output Format:
Only output the final Cedar policy inside <cedar_policy> tags.
<cedar_policy>
Cedar policy here
</cedar_policy>

USER MESSAGE
## Cedar Schema
{CEDAR_SCHEMA}

## Access-Control Requirement
{ACCESS_CONTROL_REQUIREMENT}

Write the Cedar policies now.
\end{datasetblock}

\begin{datasetblock}{Structured zero-shot policy-generation prompt}{lst:structured-prompt}
SYSTEM MESSAGE

You are a careful Cedar policy generator. Output only final Cedar code.

USER MESSAGE

ROLE

You are an Access Control Policy Engineer with expertise in:
- Cedar policy language
- Access control design
- Translating natural-language specifications into formal policies

Your responsibility is to generate correct, schema-grounded Cedar policies from specifications.

TASK

You are given two inputs:
1. A detailed natural-language access-control requirement
2. A corresponding Cedar schema

Generate a complete Cedar policy that faithfully implements the requirement and conforms to the provided schema.

GUIDELINES

- Faithfully implement all access-control requirements.
- Use only entity types, actions, attributes, and relations defined in the schema.
- Do not invent schema elements.
- Represent every constraint described in the requirement.
- Produce valid Cedar syntax.
- Treat the schema as the authoritative source of valid types and fields.
- Output no explanations, reasoning steps, or comments.

CEDAR SYNTAX CHEAT SHEET

Policy structure:

permit | forbid (
  principal [== | is | in] ...,
  action [== | in] ...,
  resource [== | is | in] ...
)
[when { <condition> }]
[unless { <condition> }];

- permit allows a request.
- forbid denies a request and overrides permit.
- when { C } applies the policy only when C is true.
- unless { C } applies the policy only when C is false.
- when and unless may appear in the same policy.
- Every policy statement ends with a semicolon.

Scope patterns:

principal == User::"alice"
action == Action::"read"
resource == File::"doc1"
principal is User
resource is Document
principal in Group::"engineering"
resource in Folder::"docs"
action in [Action::"read", Action::"write"]

Condition operators:

principal.age >= 18
resource.classification != "HighlyRestricted"
context.networkRiskScore < 20

Comparison operators:

== != < <= > >=

Logical operators:

principal.department == "eng" && !resource.isArchived
principal in Role::"Admin" || principal in Role::"Owner"

Attribute existence:

Guard an optional attribute with has before accessing it:

resource has expiryDate &&
resource.expiryDate > context.now

Collection membership:

principal in resource.readers
resource.tags.contains("public")

Attribute access:

principal.department
resource.owner
resource.project.status
context.isCompliantDevice

KEY RULES

1. Use only entity types, actions, and attributes defined in the schema.
2. Do not invent names.
3. Always guard optional attributes with has.
4. Use in for entity hierarchy or entity-set membership.
5. Use .contains() for primitive-set membership.
6. Remember that forbid overrides permit.
7. Cedar denies a request unless at least one applicable permit authorizes it.

POLICY SKELETON

permit (
  principal is User,
  action == Action::"some_action",
  resource is SomeResource
) when {
  true
};

ACCESS-CONTROL REQUIREMENT

{ACCESS_CONTROL_REQUIREMENT}

CEDAR SCHEMA

{CEDAR_SCHEMA}

OUTPUT FORMAT

Output only the final Cedar policy inside <cedar_policy> tags.

<cedar_policy>
Cedar policy here
</cedar_policy>
\end{datasetblock}

\end{document}